\documentstyle[12pt,epsf,a4]{article}

\def\inbar{\,\vrule height1.5ex width.4pt depth0pt}
\def\I1{\relax{\rm 1\kern-.33em 1}}
\def\IK{\relax{\rm I\kern-.18em K}}
\def\IQ{\relax\hbox{$\inbar\kern-.3em{\rm Q}$}}
\font\cmss=cmss10 \font\cmsss=cmss10 at 7pt
\def\IZ{\relax\ifmmode\mathchoice
{\hbox{\cmss Z\kern-.4em Z}}{\hbox{\cmss Z\kern-.4em Z}}
{\lower.9pt\hbox{\cmsss Z\kern-.4em Z}}
{\lower1.2pt\hbox{\cmsss Z\kern-.4em Z}}\else{\cmss Z\kern-.4em Z}\fi}

\let\a=\alpha
\let\b=\beta
\let\g=\gamma

\let\th=\theta

\let\l=\lambda
\let\m=\mu

\let\r=\rho

\let\vf=\varphi

\def\ovr{\over}

\def\beq{\begin{equation}}
\def\eeq{\end{equation}}
\def\bear{\begin{eqnarray}}
\def\enar{\end{eqnarray}}
\def\beit{\begin{itemize}}
\def\enit{\end{itemize}}
\def\ld#1{\partial_{#1}}

\def\+-{\underline +}
\def\1/2{{1\over 2}}
\def\nonum{\nonumber \\}
\def\pdr{\partial_r}
\def\tr{\,\rm Tr\,}
\def\dzo{{\partial}_{z_1}}
\def\dzob{{\partial}_{{\bar{ z}}_1 }}
\def\dztb{{\partial}_{{\bar{ z}}_2 }}
\def\zp{\acute {z}}
\def\dzt{{\partial}_{z_2}}
\def\zpb{\acute{\bar{ z} }}

\def\omo{\omega_1}
\def\omob{{\bar{\omega}}_1}
\def\omt{\omega_2}
\def\omtb{{\bar{\omega}}_2}
\def\zob{{\bar{z}}_1 }
\def\ztb{{\bar {z}}_2 }
\def\zthb{{\bar {z}}_3 }
\def\zfb{{\bar {z}}_4 }

\def\ln{{\rm ln}}
\def\zb{{\bar z}}

\newcommand{\sect}[1]{\section{#1} \setcounter{equation}{0}}

\def\baselinestretch{1.2}
\begin{document}
\titlepage
\begin{flushright} QMW-93/15 \\
		  hep-th/9307167 \end{flushright}
\begin{center} \bf
TOWARDS A STATISTICAL MECHANICS OF NONABELIAN VORTICES\\
\rm
\vspace{3ex}
CHINORAT KOBDAJ \footnote{Work supported by Thai Government.
			  \\ \indent e-mail:C.Kobdaj@qmw.ac.uk}\\
\vspace{1ex}
and \\ STEVEN THOMAS \footnote{ e-mail: S.Thomas@qmw.ac.uk}\\
\vspace{2ex}
{\it Department of Physics \\
Queen Mary and Westfield College \\
Mile End Road\\
London E1 \\
U.K.}\\\vspace{2ex}
ABSTRACT
\end{center}

A detailed study is presented of classical field configurations
 describing nonabelian vortices in two spatial dimensions,
when a global \( SO(3) \) symmetry is spontaneously broken to a discrete
group \( \IK \) isomorphic to the group of integers mod 4. The vortices
in this model are characterized by the nonabelian fundamental group
\( \pi_1 (SO(3)/{\IK}) \cong  \IQ_8 \), which is isomorphic to
 the group of quaternions.
We present an ansatz describing isolated vortices in
this system and prove that it is stable to perturbations. Kinematic
constraints are derived which imply that at a finite temperature,
only two species of vortices are stable to decay, due to `dissociation'.
The latter process is the nonabelian analogue of
the well known instability of charge \(|q| >1 \) abelian vortices to
 dissociation into those with charge \( |q| = 1\). The energy
of configurations  containing at maximum two vortex-antivortex pairs, is
then computed.
When the pairs are all of the same  type, we find the usual Coulombic
interaction energy as in the abelian case. When they are different, one
finds novel
interactions which are a departure from Coulomb like behavior. These results
allow one to compute the grand canonical partition function (GCPF)
for thermal
pair creation of nonabelian vortices, in the approximation where  the
fugacities for vortices of each type are small.
It is found that the vortex fugacities
depend on a real continuous  parameter \( a \) which characterize the
degeneracy
of the vacuum in the model considered. Depending on the relative sizes of
these fugacities, the vortex gas will  be dominated by one of either of
the two types mentioned above. In these
regimes, we expect the standard Kosterlitz-Thouless phase transitions to
occur, as in systems of
abelian vortices in 2-dimensions. Between these two regimes, the gas contains
 pairs of both types, so nonabelian effects will be important.
The results presented should provide a starting point from which to
investigate screening properties of the  nonabelian vortex gas.

\newpage
\sect{Introduction}

 The study of vortex defects in both two and three spatial dimensions,
 has revealed many interesting properties which
 have had a wide variety of applications in both condensed matter physics
[1], and cosmology [2]. In their guise as cosmic strings [1],
vortices in three spatial dimensions, which occur as the result of
the spontaneous breakdown of a global symmetry in the early universe,
may go some way in explaining the origin of large scale structure.
At the other extreme, it has been known for some time [3] that
vortices in 3-dimensional condensed systems such as superfluid helium IV,
play an important role in understanding the properties of these systems.

 It is well
known that vortex defects in any dimension are characterized by
\( \pi_1 (M ) \), the fundamental group of the vacuum manifold \( M \)
of the theory.
The type of vortex defects that occur in for example helium IV  are `abelian'
in
the sense that they are characterized (as we shall review later),
by an abelian fundamental group, isomorphic to the group of integers
\( \IZ \).
  In 2-spatial dimensions, abelian vortices have  particularly
simple  interactions which allows one to write down the
grand canonical partition function (GCPF) for thermal pair creation to all
orders
in the vortex fugacity in terms of the so called Coulomb gas model [4].
The pioneering work of Berezinski [5], and
Kosterlitz and Thouless [6], into their statistical mechanical properties,
showed that a gas of such vortices
underwent a novel kind of phase transition at some
critical temperature \( {\rm T}_{\rm c} \). A simple physical picture
emerged where for T \( < {\rm T}_{\rm c} \), vortices and
antivortices form a medium of bound pairs which subsequentally dissociates into
free
vortices and antivortices for  T  \( >{\rm T}_{\rm c} \).
 These results when applied to
approximately 2-dimensional  systems such as helium IV thin films,
lead Kosterlitz and Nelson [7] to predict a universal jump in the superfluid
density at \( {\rm T}_{\rm c} \), which was later experimentally verified
[8].

In this paper we shall study some properties of nonabelian vortices
in two spatial dimensions, which should eventually lead to an understanding
of their statistical mechanical properties. There is particular interest
in these kind of vortex systems, not only because they are interesting
extensions of those described above, but also because they are
known to occur in certain 2-dimensional liquid crystals [9]. In fact,
we shall consider the simplest model of spontaneous global symmetry breaking
that yields a nonabelian \( \pi_1 ( M ) \cong \IQ_8 \) where the
latter is the order 8 group associated with the field of quaternions.
Vortices described by this fundamental group are just those which
are expected to occur in so called nematic
liquid crystals [9]. As we shall see in later sections, the
interactions of nonabelian vortices are very much more complicated than
their abelian counterparts, so much so that we cannot expect to be able
to write down
the complete nonabelian vortex gas partition function to all orders in
the relevant fugacities. We will only calculate the latter to fourth order
in fugacities. Whilst this may seem very far from the all orders results
in the abelian case, it is still sufficient to investigate whether
a kind of nonabelian charge screening occurs, as we shall now explain.

Ordinary abelian charge screening
is the mechanism behind the Kosterlitz-Thouless type phase transition described
above [1]. There are a number of ways of exhibiting this screening, for example
by exploiting the map between the vortex gas partition function and that of
sine-Gordon theory in 2-dimensions [10]. Alternatively one may derive a so
called
Poisson-Boltzmann (P-B) equation [11], satisfied by the  linearly screened
potential
between a test vortex and antivortex placed in the gas. The latter equation
is derived in a perturbative small fugacity expansion, and it is clear that
in order to compute the first order screening effects one needs to know
the energy of a configuration with at least two vortices and two antivortices.
One of these vortex-antivortex pairs will play the role of test charges, whilst
the other will describe lowest order screening due to thermal pair creation.
Of course due to the abelian properties, the resulting P-B equation is
relatively simple as are its solutions [1]. One learns however,  that it is not
necessary to
have an exact expression for the partition function, to investigate screening
properties.

{}From the results given in this paper, it should be possible to
generalize the notion of linearly screened potential, and to
derive a nonabelian version
of the Poisson-Boltzmann equation [12].

The structure of the paper is as follows. In Section 2 we introduce the
2-dimensional
field theory model which describes nonabelian vortices associated with
\( \pi_1 ( M ) \cong \IQ_8 \). After a brief review
of the homotopic classification of vortices,
we present an ansatz describing isolated nonabelian vortices, and give
numerical minimum energy solutions for the Higgs
fields that define their `core' region. The properties of these solutions are
shown to guarantee finite energy. In Section 3 we present an ansatz for
multivortex configurations and write down an expression for the
 GCPF describing the thermal creation
of vortex-antivortex pairs. The energy density of all relevant configurations
up to a maximum of four vortices is calculated. Section 4 describes a
method for obtaining the explicit form of the total energy of each
of these configurations. As an illustrative example, the method
 is  used to compute the energy of a particular configuration in which
nonabelian effects are expected to be important. After a conclusion and comment
on the results, Appendix A gives details of a proof that our ansatz for
isolated nonabelian vortices is indeed one of minimum energy.
We should say at this point that  there have been a number of recent
papers concerning the subject of nonabelian vortices (e.g. ref.[13]).
However these differ from our approach in that the symmetry breaking
leading to nonabelian vortices is local. Moreover, the main
investigations of these papers concerns the quantum properties
of nonabelian vortices and their connections to systems obeying
fractional spin statistics.

\sect{Definition of the model}

In this section we shall describe in detail, a Higgs system which will
give rise to nonabelian vortices based on the discrete group \({\IQ}_{8}\)
of quaternions. As stated previously, the latter are probably
the simplest examples
of nonabelian vortices, but which also have direct physical
interest. We shall consider the Higgs model first described in [14],
where the scalar field or order parameter \( \Phi \) is in the five dimensional
representation of \( SO(3)\) , and so can be written as a traceless,
symmetric \( 3\times 3 \) matrix.
The energy  density of the model is taken to be
\beq\label{eq:2.1}
{\cal E} =    \1/2 {\rm Tr}\ \,
			 \Bigl[ g^{ab}\,
			  \left( \partial_a\Phi \right)
			  \left( \partial_b\Phi \right)
			 \Bigr]   - \,V(\Phi)
\eeq
 where the potential is chosen to be
\beq\label{eq:2.2}
	V(\Phi) \ =\
		{\l\ovr 4}\tr \Phi^4+{\l'\ovr 4}(\tr\Phi^2)^2+
{\r\ovr 3}\tr\Phi^3-\1/2\m_0^2\tr\Phi^2
\eeq
and  \( g_{ab} \) is the 2-dimensional flat euclidean metric, where
\(a,\,b\,=\, 1,\,2 \) run over 2-dimensional polar coordinates
(r, \( \theta \)  ) respectively.

Any field configuration \(\Phi\)   can be diagonalized by some orthogonal
transformation
 \( A \)
\beq\label{eq:2.4}
	A\,\Phi\, A^T \equiv\,\Phi_{diag}
		  =\,{\rm diag}( \varphi_1,\varphi_2 ,-\varphi_1-\varphi_2).
\eeq
Since  \( V(\Phi) \) is \( SO(3)\) invariant, we have
\bear\label{eq:2.5}
V(\Phi)
  &\ =\ & V(A\,\Phi\,A^{-1})\ =\ V(\vf_1,\vf_2)
      \nonum
  &\ =\ &
      {\l\ovr 4}(\varphi_1^4+\varphi_2^4+(\varphi_1+\varphi_2)^4)
     +{\r\ovr 3}(\varphi_1^3+\varphi_2^3-(\varphi_1+\varphi_2)^3)
      \nonum
  &-& \1/2\m_0^2(\varphi_1^2+\varphi_2^2+(\varphi_1+\varphi_2)^2)
      +{\l'\ovr 4}[(\varphi_1^2+\varphi_2^2+(\varphi_1+\varphi_2)^2)]^2
\enar
The extrema of the potential, are given by the solution to the equations
\bear\label{eq:2.6}
0 &\ =\ &
     \l(2\varphi_1^3+\varphi_2^3+3\varphi_1^2\vf_2+3\varphi_1\vf_2^2 )
     +\r(-\varphi_2^2-2\varphi_1\varphi_2)
      \nonum
 &-& \m_0^2(2\varphi_1+\varphi_2)
      +\l'(4\varphi_1^3+2\varphi_2^3+6\varphi_1^2\vf_2+6\varphi_1\vf_2^2)
      \nonum
\enar
\bear\label{eq:2.7}
 0 &\ =\ &
      \l(2\varphi_2^3+\varphi_1^3+3\varphi_1^2\vf_2+3\varphi_1\vf_2^2)
     +\r(-\varphi_1^2-2\varphi_1\varphi_2)
      \nonum
 &-& \m_0^2(\varphi_1+2\varphi_2)
      +\l'(2\varphi_1^3+4\varphi_2^3+6\varphi_1^2\vf_2+6\varphi_1\vf_2^2)
      \nonum
\enar
Writing \( \varphi_2 = a\,\varphi_1 \, \)for some real parameter \(a\),
we have to solve two simultaneous cubic equations
(\ref{eq:2.6}), (\ref{eq:2.7})  in say the variable
\( \varphi_1 \). For \( \rho = 0 \), but
 arbitrary \( \l ,\, \l' \) one finds that there
 are degenerate set of minima which lie on an ellipse in the
\( \varphi_1-\varphi_2  \) plane, defined by the equation
\beq\label{eq:bla}
	[(\lambda+\lambda')\,(\,\vf_1^2+\,\vf_1\,\vf_2+\,\vf_2^2 )\,
	-\,1]=0
\eeq
where in  eqn.(\ref{eq:bla}), and in the rest of the paper,
we absorb the mass scale \( \mu_0\) into the fields
\( \vf_1,\ \vf_2\).
For varying values of \( \l \) and \( \l' \), eqn.(\ref{eq:bla})
defines a continuum set of ellipses,
 illustrated in Fig.1 .   If
\( \rho  \neq 0\), then in fact this degeneracy is broken
to a large extent, and
the potential has three absolute minima at the points
\( \varphi_2 =\,a\,\varphi_1  \) on the {\it same} ellipses,
 with \( a = 1,\, -1/2 \) and \(-2\) respectively.
These minima are indicated by the intersection of the three dashed radial
lines with the ellipses in Fig.1. Since the vacuum
expectation value \( <\Phi> \) can be written in terms of
\( \varphi_1 \) and \(a\) at a minimum, one has
\beq \label{eq:2.8}
	<\Phi>
		= \varphi_1 \pmatrix { 1 & 0         &   0     \cr
			    0         & a &   0     \cr
			    0         & 0         & -1-a\cr}
\eeq
It is clear from eqn.(\ref{eq:2.8})
that when \( \rho \neq 0 \), the three absolute minima
correspond to an unbroken \( U(1) \) global symmetry, when \( <\Phi> \)
is invariant under rotations about the \(x,\, y\) and \(z\)-axes
, for \(a = 1,\, -1/2\) and \(-2\) respectively. When \( \rho =
0 \) there is again an unbroken \(U(1)\) at the above
three points as well as  at three new points corresponding
to the intersection of the ellipse with the dotted radial lines in Fig.1.
The dotted lines shown in this figure are just the continuation through the
origin of the
dashed radial lines, which together define  the lines \( \vf_2 = a\,\vf_1 \)
for the three
values of \( a\) just given. For \( \rho \neq 0 \),
the vacuum manifold \( M\), i.e.,
the set of points in the space of matrices given by
\( \Phi = {\cal G}<\Phi>{\cal G}^{-1}\), \( {\cal G} \,\in\,SO(3) \)
, will be given by the quotient
space \mbox {\( M\) = {\( SO(3)\)}/{\( U(1) \)}}.
 The fundamental group of \(M\) in this case will be isomorphic to \( \IZ_2 \),
the group of integers modulo 2. As such, the
vortices which can arise after this symmetry breaking
will necessarily be abelian in nature. Since we
wish to investigate vortices of the nonabelian
 variety in this paper, we shall take the coupling \(\rho\) always to be zero.
 Then as long as we choose values for \( <\vf_1> \) and \(<\vf_2> \)
 on the ellipse which do not
 coincide with the six points at which \( U(1)\) symmetry remains unbroken,
 the maximum symmetry group of \( <\Phi> \) will be a discrete subgroup
 \( \IK \) of
  \( SO(3) \), which is isomorphic to the group of integers mod 4. As we
 shall see in more detail below,  this leads to
 the manifold \( M \) having a fundamental group given by the nonabelian
 discrete group \( {\IQ}_8 \), isomorphic to the group of quaternions.
 Throughout the rest of this paper, we shall concentrate on this case.

We  now move on to discuss in detail an ansatz for isolated nonabelian
vortices with specific reference to the group \( {\IQ}_{8} \). In
appendix A we shall show that our ansatz is the correct one in
that it describes vortices with the lowest
energy. The vacuum manifold for the case when \( <\Phi > \)
breaks \( SO(3) \) down
to the discrete subgroup \( \IK \) which has elements
\bear \label{eq:2.9}
{\cal K}_1 &= &
\pmatrix{ 1 &  0  &  0 \cr
	  0 &  1  &  0 \cr
	  0 &  0  &  1 \cr  }
\qquad
{\cal K}_2 =
\pmatrix{ -1 &  0  &  0 \cr
	   0 & -1  &  0 \cr
	   0 &  0  &  1 \cr  }
\nonum
{\cal K}_3 & =  &
\pmatrix{ -1 &  0  &  0 \cr
	   0 &  1  &  0 \cr
	   0 &  0  & -1 \cr  }
\qquad
{\cal K}_4 =
\pmatrix{  1 &  0  &  0 \cr
	   0 & -1  &  0 \cr
	   0 &  0  & -1 \cr  }
\enar
is given by \( M = \ SO(3)/\IK \cong SU(2)/\IQ_8  \),
where we have used the well known
homomorphism between \( SO(3) \) and \( SU(2) \). It follows that
vortices are characterized by \( \pi_1 (M ) \cong \IQ_8 \). Using the
homomorphism between \(SO(3)\) and \( SU(2) \) we can equally work
with fields \( \Phi \) which are representations of either group. We will
usually work with \( SO(3) \) representations and group elements
(the latter denoted by \( \cal{G} \) ) but will often give the corresponding
\( SU(2) \) fields and elements (denoted by \(g\) ), on which \( \IQ_8 \)
has a simple action.

Our ansatz for a single vortex  at the origin
in the \( (r, \theta ) \)
plane will be
\beq\label{eq:2.10}
	\Phi(r,\theta) \ = \ {\cal G }(\theta)\, \Phi_{diag}(r) \,
				\,{\cal G}^{-1}(\theta)
\eeq
where in eqn.(\ref{eq:2.10}), \( \Phi_{diag} \) denotes the diagonal matrix
\mbox{(\( \varphi_1 (r), \varphi_2 (r),
- ( \varphi_1 (r)+\varphi_2 (r) ) \) ).}
It is assumed that \( \varphi_1 , \varphi_2 \) are functions
of \(r\) only whilst the group elements \( \cal{G} \) depend only
on \( \theta \). It is important that on encircling the vortex
at \(r = 0 \), i.e. as \( \theta \) is varied from 0 to 2\(\pi \),
\( \Phi (r, \theta ) \) be strictly single-valued.
The  corresponding boundary conditions on \( \cal{G} (\theta ) \)
consistent with this constraint are,

\beq\label{eq:2.11}
      {\cal G}(\theta + 2\,\pi )\, = \,{\cal G}(\theta )\, h
\eeq
where h is an element of \( \IK \). It might appear that there are
only four `types' of vortices corresponding to the four elements of \( \IK \).
However it should be remembered that \( SO(3) \) is not simply
connected since \( \pi_1 (SO(3)) \cong \IZ_2 \), which
effectively doubles the number of homotopically
inequivalent closed curves on \( M \). This is more
transparent when working with \( SU(2) \) group elements.
Under the homomorphism \( SU(2) \cong SO(3) \),
the group of quaternions \(\IQ_8 \) gets mapped to the
abelian group \( \IK \). So somewhat hidden in the
boundary conditions given in
eqn.(\ref{eq:2.11})  is the fact that closed curves on \(M\)
are actually characterized by elements of \( \IQ_8 \).

Next we need find out what are the explicit representations
of \( \pi_1 (M ) \). From a mathematical point of view there
are of course, an infinite number of homotopically equivalent
 representations. However, we are only interested in
the physically relevant ones, i.e. those which yield a minimum
energy for the corresponding vortex, and we expect that such
 `minimal' representations will be unique.
The elements \( {\cal K}_2 ,..., {\cal K}_4 \) given in eqn.(\ref{eq:2.9})
correspond to discrete rotations by 180 degrees about the \(x, y\) and
\(z\)-axes respectively. Therefore, a simple ansatz for the
\( SO(3) \) elements \( \cal{G} \) which satisfy the boundary conditions
in eqn.(\ref{eq:2.11}) for a given element of \( \IK \), is for them to be
rotations by angles \( \theta /2 \) about the respective axes.
 In fact, we shall show in Appendix A that
this ansatz actually corresponds to minimal energy vortices.

In Table 1, we have listed these elements, together with the
corresponding \( SU(2) \) representations. Also listed are the
discrete group elements that appear in the boundary conditions
as we encircle the vortex. The \( SU(2) \) representations
in particular,
make it clear that this group is \( \IQ_8 \). We have used the
notation  \( {\cal G}_i , {\cal G}_j , {\cal G}_k \),
\({\cal G}_i^{-1}, {\cal G}_j^{-1}, {\cal G}_k^{-1}\) and
\( {\cal G}_{-\I1} \) to
denote group elements whose boundary conditions are twisted
by the elements \(i, j, k, -i, -j, -k \) and -\I1
 of \( \IQ_8 \), where
\bear\label{eq:temp}
 {i}^2
= {j}^2 = { k}^2 = -\I1
\nonum
i\, j \,= \, k \quad ,\quad  j \, k \, = \, i \quad,\quad k\, i\, = \,j
\enar
 are the
multiplication rules of \( \IQ_8 \). The parameter \({\theta}_{0} \)
in Table 1 is an arbitrary constant associated with the origin
of the \(\theta\) coordinate.

Given these explicit forms for \( {\cal G} \), we want to minimize
the energy in eqn.(\ref{eq:2.1}) in the remaining fields \( \varphi_1 (r) ,
\varphi_2 (r) \), in order to find stable vortex solutions.
Because of the \( \Phi \rightarrow -\Phi \) symmetry of the
energy, (assuming \( \rho = 0 \) ), the equations
for minimizing the energy will be the same for \( {\cal G}_{\alpha} \)
as for \( {\cal G}_{\alpha}^{-1} \), for  \( \alpha \  \epsilon \ {\IQ}_8 \).
We shall  refer to these as being vortices and antivortices of
type \(\alpha \) respectively. Furthermore, the
vortices of type -\I1 are by nature abelian since -\I1 commutes
with all other elements in \( \IQ_8 \). Moreover, we shall see in the next
section that in a finite temperature
system they are unstable to dissociation into either \( i, j \) or \( k \)
type vortex-antivortex pairs. For these reasons we
will concentrate for the remainder of the paper,
on vortices of type \(i, j\) and \(k\). The parameter \( \alpha \) which
denotes species type, will therefore implicitly run over only these
particular types.

Substituting the ansatz for \( {\cal G}_k \) in these three cases
into the energy we get
\bear\label{eq:2.12}
E_{\alpha}
&=& \int d^2 x\ \Biggl\{  \1/2 {\rm Tr}\ \,
			 \Biggl[ \,g^{ab}   \,
			  \left( \partial_a\Phi \right)
			  \left( \partial_b\Phi \right)
			 \Biggr]   - \,V
		\Biggr\}\nonum
&=&
	\int_0^\infty dr\, r \int_0^{2\pi} d\theta\ \,
	\Biggl\{ \1/2 (\pdr \vf_1)^2
		    +\1/2(\pdr \vf_2)^2+\1/2 (\pdr(\vf_1+\vf_2))^2 \nonum
	  &&\qquad + \ {1\ovr {2\,r^2}} \,{ P}_{\alpha} (\vf_1 , \vf_2 )
		    - V(\vf_1,\vf_2)
	\Biggr\}
\enar
The effect of the terms \( {\cal G}_{\alpha} (\theta ) \)
on the energy \({ E}_{\alpha} \)
 is to add certain potential terms, denoted by
\( {1\ovr{r^2}}\,{ P}_{\alpha } (\vf_1 , \vf_2) \)
in eqn.(\ref{eq:2.12} ) which  are given by
\beq \label{eq:2.13}
     { P}_i \ = \   {1\ovr 2 }\,(\vf_1\,+\,2\,\vf_2)^2
\eeq
\beq
     { P}_j \ = \   {1\ovr 2 }\,(2\,\vf_1\,+\,\vf_2)^2
\eeq
\beq
     { P}_k \ = \    {1\ovr 2 }\,(\,\vf_1\,-\,\vf_2)^2
\eeq
The minimization equations for \( { E}_{\alpha } \) are
\def\pd{ \partial  }
\bear\label{eq:2.14}
    {2\ld{rr}\vf_1 +{2\ovr r}\,\ld{r}\vf_1}+\ld{rr}\vf_2 +{1\ovr r}
    \ld{r}\vf_2-{1\ovr {2\, r^2}} \frac{\pd {P}_{\alpha}}{\pd\vf_1}
 + \frac{\pd V}{\pd\vf_1} &= &0
\nonum
    {2\ld{rr}\vf_2 +{2\ovr r}\,\ld{r}\vf_2}+\ld{rr}\vf_1 +{1\ovr r}
    \ld{r}\vf_1-{1\ovr {2\, r^2}} \frac{\pd { P}_{\alpha}}
{\pd\vf_2} + \frac{\pd V}{\pd\vf_2}&=&0
\enar
On physical grounds, (namely the finiteness of the core energy
\( { E}_{\alpha}^{c} \), which we will be defined later),
we anticipate that the solutions to these equations that minimize
the energy have the following asymptotic
behavior in the radial coordinate \(r\). As \( r \rightarrow 0 \),
it is sufficient that
\( \vf_1 \) and \( \vf_2 \) both vanish at least as fast as \(r\),
whilst for
 \( r \rightarrow \infty \), we expect both
\( \vf_1 \) and \( \vf_2 \) go to their vacuum values, which
minimize \(V \). To show that smooth interpolating functions exist
with these asymptotic properties, one may solve the non-linear
differential equations (\ref{eq:2.14}) numerically. This was done by
discretizing the energy in eqn.(\ref{eq:2.1}) and carrying out
numerical minimization, for various values of the parameters
\( \l, \l'\) and vacuum parameter \( a \). Fig.2. illustrates the solutions
found in the case of an \(i\)-type vortex with
\( \l = 0.01 \) and 2,  \( \l' = 0 \) and \( a= -1 \) (we shall
explain why we are interested in the value \(a\) = -1 later).
In Fig.2,
the dotted curves show the numerical solutions whilst the
solid ones are those of a trial solution, needed as input
in the minimization procedure. These trial solutions are
of the form
\bear\label{eq:2.15}
  \vf_1 |_{trial} & = & \frac{1}{ \sqrt{\lambda( 1+a+a^2)} }
			\left ( \frac{r\mu_0}{r\mu_0 +1 } \right )\nonum
  \vf_2 |_{trial} & = & \frac{a}{ \sqrt{\lambda( 1+a+a^2)} }
			\left ( \frac{r\mu_0}{r\mu_0+1}   \right )
\enar
The normalization factors in eqns.(\ref{eq:2.15})
 are such that \(\vf_1 (\infty ),
\vf_2 (\infty ) \) lie on the ellipse of degenerate vacua,
described in the previous section.

 As one can verify from Fig.2, these trial solutions are very good
approximations, particularly close to the origin of \( r \). It
is also apparent that there is a `core' region
(marked as \( r\) = \( {r}_{c} \) in Fig.2) beyond
which the values of \( \vf_1 , \vf_2 \) do not change appreciably,
from their vacuum values \footnote{ Although there is no definite value to
\(r_c\), we can define it to be that value of \(r \) such that
\mbox{\( \vf_1/<\vf_1>=\vf_2/<\vf_2>=1-{1\ovr e} \)}, that is
\(\vf_1,\,\vf_2\)
are within \({1\ovr e }\) of their vacuum values. This gives
\(\mu_0 r_c \sim \, 1.72\) }.
Of course, different definitions of \(r_c\) are possible, depending on how
close to unity one requires the ratios \(\vf_1/<\vf_1> \) and
\(\vf_2/<\vf_2>\) to be. The lack of a precise core radius is not a particular
property of nonabelian vortices, but is also apparent in the abelian case.
 What is clear in both cases however, is that different choices for \(r_c\)
will simply give rise to constant factors in the expression for the self
energy of a vortex (see eqn.(\ref{eq:2.16}) below).
Similar solutions to these exist for
\(j\) and \(k\) type vortices, the only difference is in the algebraic
form of the potentials \( { P}_{\alpha} \) which, it turns
out, has a minimal effect on the behaviour of the solutions.
 It will
be a very useful (and good) approximation to replace the numerical
solutions for \( \vf_1 , \vf_2 \) of \(i,\,j\) and \(k\) type vortices
by those  of the trial solutions given in eqns.(\ref{eq:2.15}) for
the region \(r \le { r}_{c} \), and by \( < \vf_1 >, < \vf_2 >
\) for \( r  >  { r}_{c} \).

This being the case, it is straight forward to compute the energy of
a single \( i,\  j \) or \( k \) type vortex (or antivortex), placed in a
2-dimensional circular box of radius \( R \). The result is

\beq\label{eq:2.16}
    { E}_{\alpha}
     \ = \  { E}_{\alpha}^{c} + \pi\,< {P}_{\alpha}>
	      {\ln }\,\left( { R \ovr {{ r}_c }}\right)
	      \quad \quad \alpha = { i,j,k}
\eeq
where  \( { r}_c \) is the core radius of
the corresponding vortex configuration, and is typically of order
unity in units of inverse mass \( {\mu }_0^{-1} \).
The logarithmic self energy terms in eqn.(\ref{eq:2.16}), correspond to
contributions coming from the region \( r > { r}_{c} \).
The core energy,
\( { E}_{\alpha}^{c} \) in eqn.(\ref{eq:2.16}) can
be approximated by taking the range of integration in eqn.(\ref{eq:2.12})
to be between \( r = 0 \) and \( r = {r}_{c}  \), and is given by
\bear\label{eq:2.17}
	{ E}_{\alpha}^{c}
	&=&  \pi\,< {P}_{\alpha}>
		\int_{0}^{r_c} \frac{ {\rm d}r}{r}
		\left ( \frac{r}{r+1} \right)^2 \,
		+\, f(<\vf_1>,<\vf_2>,r_c)
		\nonum
	&=& \pi\,< {P}_{\alpha}> \{ {\ln}\, (r_c +1 )
	      + \frac{1}{r_c +1} -1 \} \,
	      +\, f(<\vf_1>,<\vf_2>,r_c)
\enar
The first term in eqn.(\ref{eq:2.17}) comes from the potential \(P_{\alpha}\)
present in \(E_{\alpha}\) due to vortices of type \(\alpha\).
The second term \(f(<\vf_1>,<\vf_2>,r_c)\) represents contributions
coming from other terms in (\ref{eq:2.12}), which are independent
of the vortex type \(\alpha\). These can be easily calculated, but
explicit expressions will not be needed in what follows.
It is worth comparing the results of eqns.(\ref{eq:2.16}), (\ref{eq:2.17})
 to those one would
obtain for a single  abelian  vortex in 2-dimensions, such as
those produced in spontaneous breaking of \( U(1) \)
symmetry [15]. In the abelian case we would also obtain an energy
formula similar to eqn.(\ref{eq:2.16}),
with the coefficient of the logarithm term proportional
to \(q^2\,<\phi^2>\), where \(q\) is the vorticity and
\(<\phi>\) is the \(U(1)\) breaking vacuum expectation value.
Hence the factors  of  \( < {P}_{\alpha } > \)
in eqn.(\ref{eq:2.16}), (\ref{eq:2.17}) for fixed \(<\vf_1>\) and
\(<\vf_2>\) are proportional to a kind of nonabelian
vorticity or nonabelian charge. The difference between the
 abelian and nonabelian example is that in the former, we can have
arbitrarily large (but quantized) vorticity \(q\), whilst
in the latter there are clearly only a finite number of
possibilities, given by the range of \(\alpha\). This is a consequence of
\( {\pi }_1 (M ) \cong \IZ  \), ( \( \IZ \)
being the group of integers), and  \( {\pi }_1 (M )
\cong {\IQ}_8 \) in the abelian and nonabelian cases respectively.

Finally, we can understand the logarithmic behavior in eqn.(\ref{eq:2.17})
and in particular
why there is a logarithmic divergence as \(R\rightarrow \infty \),
even for nonabelian vortices. This is because,
as the numerical results suggest, one can think of these vortices as
having a definite core region, which for distances  \( r \gg
{r}_c \), appears approximately as a point-like charge. It is
natural then that the field energy outside such an isolated point charge be
divergent as we go to the infinite volume limit, whilst the logarithmic
behaviour is simply a consequence of the   2-dimensionality of the system. In
fact,
to see the truly novel effects of nonabelian vortices we have
to study more than simply self-energies. Clearly we have to
consider multi-vortex configurations where at least two of the
vortices are of types h and g where  the latter are two non-commuting
elements of  \(\IQ_8 \).
We shall study this situation in the next section .

\sect{Systems of multi-vortices}

Having discussed at length in the previous section the ansatz corresponding
to isolated nonabelian vortices we now move on to discuss multiple
vortex configurations. The simplest such configuration would be a
vortex-antivortex
pair of type \( \alpha \). Since we are ultimately interested in
the statistical mechanical properties of such vortices, one
should consider what happens  in
a real system at finite temperature.
The logarithmically divergent self energy of a single vortex discussed in the
last section, prohibits, in the thermodynamic limit,
 single vortex creation in these systems
due to thermal fluctuations. This is a property which is familiar in  the case
of abelian vortices [1].  However, neutral vortex-antivortex pair
creation is not suppressed in this way, and one should
therefore expect these to be present  at  finite temperature.

Intuitively we expect that whatever the correct description of such a pair is,
it should go over to the previous ansatzes for an isolated vortex and
antivortex
in the limit where we take the distance between the pair very large with
respect
to their typical core size \(r_c \). As such we shall take as our ansatz
for a vortex of type \(\alpha \) at the point \( z_1 \) and an antivortex
of the same type at \(z_2\) to be
\beq\label{eq:3.1}
	{\cal G}_{\alpha}(z,z_1,z_2) \simeq {\cal G}_{\alpha}(z,z_1)
	{\cal G}_{-\alpha}(z,z_2)
\eeq
where in eqn.(\ref{eq:3.1}), the \({\cal G}_{\alpha} \) correspond to the
1-vortex solutions obtained in the previous section.
The corresponding field configuration denoted by
\( \Phi_{\alpha}\) is
\beq\label{eq:3.2}
  \Phi_{\alpha}\,=\,{\cal G}_{\alpha}(z, z_1) \,{\cal G}_{\alpha}^{-1}(z, z_2)
		   \Phi_{diag}(z, z_1, z_2)
		       \,{\cal G}_{\alpha}(z, z_2)\,
		       {\cal G}_{\alpha}^{-1}(z, z_1)
\eeq
Again, for well separated pairs
we expect that the `core' field \( \Phi_{diag} \)  vanishes
 as \(z\) approaches either of the vortex centre's at \( z_1 \) or \( z_2 \) ,
 (which  ensures
a finite core energy in each case) just as in the 1-vortex case.
In fact this behavior can be confirmed if one were to solve the
minimum energy equations for \( \vf_1, \vf_2 \) for the vortex pair
with \( {\cal G}_{\alpha} \) given by eqn.(\ref{eq:3.2}). As
\( z \)
approaches either of \(z_1 \) or \( z_2 \), the form of
\({\cal G}_{\alpha} \) goes over to that of the single vortex,
 modulo an unimportant constant shift in the angular
coordinate \( \theta \). Hence from our results of the last section,
it is clear that the solutions for \(\vf_1 , \vf_2 \)
will become, in the limit of large pair separation, those
described in the previous section. Thus as before we can approximate
\( \vf_1 ,\vf_2 \) to be given by their vacuum values at
distances \( | z - z_s | \geq   r_c \),
where \(s\) = 1, 2 . For distances \( |z - z_s | < r_c \) we
can approximate the fields by the trial solutions
given in eqn.(\ref{eq:2.15}). Putting all this together we
find that the energy of a  vortex-antivortex pair is given by

\beq\label{eq:3.3}
   E_{\alpha}(z_1,z_2)\, = \,2 E_{\alpha}^{c} + 2 E_{\alpha}^{self} +
				E_{\alpha}^{int} (z_1 , z_2 )
\eeq
where in eqn.(\ref{eq:3.3}) the core energy \(E_{\alpha}^{c} \) was given
in (\ref{eq:2.17}). The self
energy \(E_{\alpha}^{self} \) and interaction energy \( E_{\alpha}^{int} \)
are computed to be
\bear\label{eq:3.4}
E_{\alpha}^{self}
	& = & \pi\, < {P}_{\alpha}> \ln ({R \ovr {r_c}} )\nonum
E_{\alpha}^{int}(z_1 , z_2)
	&=&
   2\,\pi\,< {P}_{\alpha}> \ln |\frac{z_1-z_2}{R} |
\enar
where \(R\)  is the size of the system, and the functions \( {P}_{\alpha} \)
were defined in eqns.(\ref{eq:2.13})-(2.17). Note that we can take the
thermodynamic
limit \( R \rightarrow \infty \) in eqn.(\ref{eq:3.4} )
since \( R \) cancels between the self and interaction energies. This is simply
a consequence of taking a vortex -antivortex pair-which can be thought
of as neutral\footnote{ In the nonabelian case, this means that
\( {\cal G}(z, z_1,z_2 ) \)is single valued upon simultaneously encircling
the points \( z_1 \) and \( z_2 \).}.
 This phenomenon is already known to occur in the
case of abelian vortices, and indeed \( E_{\alpha}^{int}
(z_1,z_2 ) \) above, is
just the standard Coulomb type of interaction which one also finds in this
 case. This is not surprising because simply studying
a vortex-antivortex
pair of type \( \alpha \) does not probe  the nonabelian structure of the
theory since \({\cal G}_{\alpha} \) and \( {\cal G}_{-\alpha } \)
are commuting elements. Clearly to study the effects due to
non-commuting elements \( {\cal G}_{\alpha} , {\cal G}_{\beta } \),
we have to look at least to configurations containing four vortices:
a vortex-antivortex pair of type \( \alpha \) interacting with a pair
of different type \( \beta \).

 Before studying the details of this, we can
make life easier by consideration of what one might call
kinematical constraints associated with the thermal creation
of vortices. As we shall see this implies that for any given value
of the vacuum parameter \( a \), only two types of vortices will be
stable to dissociation into one another, where these types depend on
the specific value of \( a \).

Let us imagine creating vortex-antivortex pairs in a system by
gradually increasing the temperature.  The self energy
eqn.(\ref{eq:3.4}) of a vortex of type \( \alpha\) in the pair is
proportional to the quantity \(<P_{\alpha}>\) (see eqn.(\ref{eq:3.4}).
As we discussed earlier,
the \(<P_{\alpha}>\)  factors can be thought of as being proportional to
a generalization of
vortex charge or vorticity to the nonabelian case. As such
one has to consider whether a nonabelian vortex of given charge is unstable
to `splitting' into two other vortices in such a way as to
conserve overall charge, yet lower the self energy.

 To understand this process let us consider the abelian case once more.
Here the vortex  self energies
are proportional to \( q^2 \) , where \(q\) is the quantized
vortex charge. Thus it is clear that a charge \(q\) = \(\pm \)2 vortex will
be unstable to dissociation into two charge \( \pm 1\) vortices
because \( E_{q={\pm 2}}^{self} > 2 E_{q={\pm 1}}^{self} \). The same is
true for all higher charged abelian vortices with \( | q |> 1 \), so that
in this sense \( |q| = 1 \) vortices are ultimately stable [15]. This
explains why only \(q\) = 1 vortices are considered in the context of
Coulomb gas models and K-T type phase transitions [1].
We want  to apply a similar analysis in the nonabelian case.
The main differences are that we have a finite number of
possible charges in this case as explained earlier, and the rules for
possible dissociation of vortices are governed by the group relations of
\( \IQ_8 \) not simply by the addition of charges.

It should be stressed at this point that the stability to dissociation
we are discussing is different to, and must be considered in addition to,
the perturbative stability analysis performed in Appendix A. Clearly
the process of dissociation of a vortex is non-local, and is separate from
the stability of a vortex  to local perturbations.

To begin with consider the self energy of a -\I1 type vortex. We saw in the
previous section that there were three kinds of ansatz for the \( SO(3)\)
element,
denoted by \( {\cal G}_{-\I1}^{(\zeta)}\), where \( \zeta \)
runs over the elements \(i, j, k\) of \( \IQ_8 \)
(see Table 1). It is easy to see that the corresponding self
energies \( E_{-\I1}^{self, (\zeta)} \)satisfy the relations
\beq\label{eq:3.5}
E_{-\I1}^{self,(\zeta)} \, = \, 4 E_{\zeta }^{self} > 2 E_{\zeta}^{self}
\eeq
for arbitrary values of \( a \). It is clear that -\I1 type vortices
are always unstable to dissociation
into a pair of vortices of type \(i, j\) or \( k \).
Which actual decay branch occurs  depends on the specific value
of \( a \) and will be whatever is the lowest energy of the three
possibilities
\( E_{-\I1}^{self,(\zeta)} \). From a practical point of view however, this
result means that we need not consider -\I1 type vortices at all in
investigating the statistical mechanics of nonabelian vortices. They are
like the \( |q| > 1 \) abelian vortices referred to above.
Let us now consider
the remaining cases of type \( \alpha\) (clearly since \( E_{\alpha}^{self}
 = E_{-\alpha}^{self} \) the arguments will hold equally for antivortices ).

Using the group composition rules for \( \IQ_8 \), a \( k \) type vortex for
example, can dissociate into an \(i \) and \(j \) type (or -\(i\)
and -\(j\) type ) if
\bear\label{eq:3.6}
E_{k}^{self} \,& >& \, E_{i}^{self} + E_{j}^{self} \nonum
	&\Rightarrow & {(1 - a )}^2 \, > \, {(1 + 2 a ) }^2
	+ {(a + 2 )}^2  \qquad  {\rm i.e. }
	-2 \, < \, a \, < - \frac{1}{2}
\enar
where in eqn.(\ref{eq:3.6}), we have substituted  for \( \vf_2 \)
in terms of the parameter \( a \). Thus in the interval \( -2 < a < - 1/2 \) ,
referred to as region \( {\rm{\bf I}}_{ij} \) in Fig.3
, \( k \) -type  vortices dissociate, and at
most we would expect to just find vortices and antivortices
of type \(i\) and \(j \) in this region. At the same time we learn
that \(i \) and \( j \) -type
are  stable to further dissociation, since in either case the
path (restricted  by the composition rules of \( \IQ_8 \)),
necessarily involves producing a \(k\) type vortex. Notice that while
\({\rm{\bf I}}_{ij} \)   is a connected portion of the real
line of \( a \) values, the corresponding   regions in
the space of vacuum expectation values
\( <\vf_1 > , <\vf_2>  \) are two disconnected pieces.
   A similar analysis of
possible decays of \(i \) and \( j \)  types  for \( a \) values
outside of region \({\rm{\bf I}}_{ij} \) gives the following results,
\bear\label{eq:3.7}
E_{i}^{self} \,&>& \, E_{j}^{self} + E_{k}^{self} \nonum
	    &\Rightarrow& \, \, {(1+2a)}^2 \, > \, {(2 + a )}^2
	    + {(1 - a )}^2  \qquad  {\rm i.e.} \,\, a\,<\,-2 \ {\rm or}\
		a\,>\,1
\enar
and finally
\bear\label{eq:3.8}
E_{j}^{self} \, &>& \,  E_{k}^{self} + E_{i}^{self} \nonum
	     & \Rightarrow & \ {(2+a)}^2 \, > \, {(1+2\,a )}^2
	     + {(1-a)}^2 \qquad {\rm i.e.} \,  -\frac{1}{2} < a < 1
\enar
the different regions of \( a \) given in eqns.(\ref{eq:3.7}, \ref{eq:3.8} )
are denoted by \({\rm{\bf I}}_{jk} \) and \( {\rm{\bf I}}_{ki} \)
in Fig.3.  Fig.3
clearly shows that the three regions fit together exactly, to
cover the entire ellipse of vacuum states, without overlap. This means that
the two types of vortices indicated by the subscripts on \( {\rm{\bf I}}_{ij},
{\rm{\bf I}}_{jk} \) and \( {\rm{\bf I}}_{ki} \) are stable
to further dissociation.
It is interesting to see that all  regions are separated by points
with either \( a = -1/2, 1 \) or \(2 \), which correspond to an  unbroken
\( U(1)\) symmetry (see Section 2 ). At such points there are no vortices,
so it appears that one cannot smoothly go from one region to another
by varying \( a\). In fact  the relative positions of the
regions \({\rm{\bf I}}_{ij},
{\rm{\bf I}}_{jk} \) and \({\rm{\bf I}}_{ki} \) in Fig.1. can be understood
in the following way.

The point where any two regions touch  has an unbroken \( U(1)\)
symmetry that
correspond to rotations about either the \(x, y \) or \(z\) axis, depending on
which
of the six  points,
where regions \({\rm{\bf I}}_{ij} \) etc meet, we consider. For example at the
point where
 \( {\rm{\bf I}}_{ij} \) and \( {\rm{\bf I}}_{jk} \) meet, the \( U(1)\)
symmetry corresponds to
rotations about the \( y \)-axis (which one can easily see from the form of
\( <\Phi > \) (see eqn.(\ref{eq:2.8})) at this point).
In some sense, we can now understand why this
particular symmetry should appear here because
 in `passing' from  \( {\rm{\bf I}}_{ij} \) to \( {\rm{\bf I}}_{jk}\),
 the vortices of type \( i \)
have been rotated by 180 degrees about the \( y\)-axes into those of type
\( k \) whilst leaving those of type  \( j \) invariant.
The other regions in Fig.3,
and their relation to the particular \( U(1)\) symmetry present at their
intercept, can be similarly understood.

The end result of the above kinematic considerations, is that for any given
ratio  \mbox{\( <\vf_2 > /< \vf_1 >  = a \)}, we need only concern ourselves
with
the statistical mechanics of two species of vortices. Let us for definiteness
choose \( a \) to be in the region \mbox{\( -2 < a < - 1/2 \)}, i.e. we only
study \( i\) and \( j \) type vortices.

 Within this approximation, we can write down a general expression for the
grand canonical partition function (GCPF) \( Z \) of thermal pair
creation of the above species of nonabelian vortices,
\bear\label{eq:3.8.5}
Z ( {\sl z}_{i},{\sl z}_{j} )
	\,& = &\,\mathop
	{\sum}_{{N}_{i} = 0 }^{\infty}
	\mathop {\sum}_{{N}_{j} = 0}^{\infty }
	\, \frac{ { {\sl z}_{i} }^{N_{i} } }{{N_{i} !} }
	\, \frac{ {{\sl z}_{j}}^{N_{j} } }{{N_{j} !} }
	\mathop {\prod}_{{\nu}_{i} = 1 }^{N_{i}}
	\mathop{\prod}_{{\nu}_{j} = 1 }^{N_{j}}
	\mathop {\int} \frac{ {\rm d}^2 {z_{{\nu}_{i}} }} {\pi {r_c}^2 }
	\mathop{\int} \frac{{\rm d}^2 {z_{{\nu}_{j}} }} {\pi {r_c}^2 } \nonum
	&\times &
	\sum_{\scriptstyle \left\{ {\cal G}_{{\nu}_{i}} \right\} \,\atop
	      \scriptstyle\,{\scriptstyle\prod\atop\scriptstyle{{\nu}_{i}}}
		   {\cal G}_{{\nu}_{i} } = {\I1 } }
	\, \sum_{\scriptstyle\left\{ {\cal G}_{{\nu}_{j}} \right\}\,\atop
		 \scriptstyle\,{\scriptstyle\prod\atop\scriptstyle{{\nu}_{j}}}
		   {\cal G}_{{\nu}_{j} } = {\I1} }
	\, {\rm exp }
	\left\{ - \frac{1}{\rm T} {\rm E}( N_{i}, N_{j},
	 {\cal G}_{{\nu}_{i}} ,
	 {\cal G}_{{\nu}_{j}} ,
	 z_{{\nu}_{i} }, z_{{\nu}_{j}} ) \right\}
\enar
In eqn.(\ref{eq:3.8.5} ), the fugacities \( {\sl z}_{i},
{\sl z}_{j} \) associated with vortices of type \( i \) and
\( j \),
are defined in terms of the core energies \({E}_{\alpha}^{c} \)
given in eqn.(\ref{eq:2.17})
\bear{\label{eq:3.8.6}}
{\sl z}_{i} & = &  {\rm exp} (- E_{i}^{c}/{\rm T}) ) \nonum
{\sl z}_{j} & = &  {\rm exp} (-  E_{j}^{c}/{\rm T})
\enar
 respectively.
 \( {\cal G}_{{\nu}_{i}} \) and \( {\cal G}_{{\nu }_{j}} \) denote the
\( SO(3) \) group elements corresponding to all the  \( i \) and \( j \) type
vortices of
a particular configuration having total energy \(E\).
 As we have already seen in this section,
only configurations with
\({\mathop {\prod}_{{\nu}_i = 1 }^{N_{i}}  {\cal G}_{{\nu}_{i}} = {\I1} }  \)
and
\({\mathop{\prod}_{{\nu}_j}^{N_{j}} {\cal G}_{{\nu}_{j}}  = {\I1 }}\)
will contribute to the
GCPF above, in the thermodynamic limit. Although eqn.(\ref{eq:3.8.5}) is
a complete formula for the GCPF, we want to study it perturbatively
in a small \({\sl z}_i\)  and \( {\sl z}_j \)
expansion since we do not know how to compute
\(E \) for arbitrary configurations of \( i\) and \( j\) type vortices.

Having considered in some detail the case of single vortex -antivortex
 pairs, we now move on to consider the next order terms
  (in a double expansion
of \( Z \)  in the fugacities \({\sl z}_i , {\sl z}_j \) ). These will be
configurations containing four vortices and antivortices arranged in two pairs,
 because the  overall neutrality
condition mentioned previously,
demands that we consider only even numbers of vortices and antivortices.
The simplest
configuration of this type consists of two vortex-antivortex pairs
of the same type \( i \) or \( j \). Because the pairs are of the
same type, they commute with each other, so  we expect that the resulting
energy be
that of the corresponding Coulomb system of charges [1,4].
The ansatz for
this system will be just the generalization of the single pair
ansatz given in eqn.(\ref{eq:3.1}), which again we expect to be a reasonable
approximation at inter-pair distances greater than the core
size \( r_c \). Thus with the understanding that the index
\( \alpha \) now runs over \( i \) and \( j \) types only, the
 \( SO(3) \) group element, \({\cal G}_{\alpha, \alpha} \),
 corresponding to  vortices of type
\( \alpha \) at points \(z_1 \) and  \( z_3 \) and antivortices
of the same type at \( z_2 , z_4 \) will be

\beq\label{eq:3.9}
	{\cal G}_{\alpha,\alpha}(z_1,\cdots,z_4) \simeq {\cal G}_{\alpha}(z,z_1)\,
{\cal G}_{\alpha}(z,z_2)\,
					{\cal G}_{\alpha}(z,z_3)\, {\cal G}_{\alpha}(z,z_4)
\eeq
with the corresponding form of \( \Phi \), which we denote by
 \( {\Phi}_{\alpha, \alpha} \) being
\bear\label{eq:3.10}
\Phi_{\alpha\, ,\alpha}(z, z_1,\cdots,z_4)
	  &=&   \ \ \,  {\cal G}_{\alpha}(z, z_1)
			\,{\cal G}_{\alpha}^{-1}(z, z_2)
			\,{\cal G}_{\alpha}(z, z_3)
			\,{\cal G}_{\alpha}^{-1}(z, z_4)
	  \nonum
	  &&    \qquad  \times
			\Phi_{diag}(z, z_1, z_2 , z_3, z_4 )
	  \nonum
	  &&    \times
			{\cal G}_{\alpha}(z, z_4)
			\,{\cal G}_{\alpha}^{-1}(z, z_3)
			\,{\cal G}_{\alpha}(z, z_2)
			\,{\cal G}_{\alpha}^{-1}(z, z_1)
			\quad;\,\alpha \ = \,i ,\, j \nonum
\enar

Again for the purposes of computing the energy of the configuration
given in eqn.(\ref{eq:3.10}), the core field
\( \Phi_{diag}( z, z_1,z_2, z_3, z_4 ) \)
will be approximated by the usual trial
functions of eqn.(\ref{eq:2.15}) whenever \( |z - z_t | < r_c \,  \)
for \( t = 1,...,4 \) or by \( <\Phi_{diag}> \) when the argument \( z \)
lies outside this region. The energy of this configuration
\( E_{\alpha , \alpha } \) is given by
\beq\label{eq:3.11}
E_{\alpha , \alpha } (z_1, z_2 , z_3 , z_4 ) \, = \,
	  4\,{E_{\alpha}^{c} } +  \pi\,< {P_{\alpha} } >
	  \sum_{{\scriptstyle s,t=1} \atop{\scriptstyle s \neq t}}^{4}  {q_s}{q_t}
\,\ln\,
	  ( \frac{|z_s - z_t |}{{r_c}} )
\eeq
where in eqn.(\ref{eq:3.11}) the charges \( q_s \) are given by
\( q_1 = q_3 = 1, q_2 = q_4 = -1 \). We see that the energy
\( E_{\alpha, \alpha }\) of two pairs of the same type \( \alpha \)
has the usual Coulombic form. Moreover, in the GCPF eqn.(\ref{eq:3.8.5}),
we have to sum over similar configurations to those above but with
different charge assignments \( q_s \) at the points \( z_s \).
However, the relative energy of these different configurations
is trivially related to that given in eqn.(\ref{eq:3.11} )
since it simply involves a permutation of the charge assignments.
This is a manifestation of the property that all the \(SO(3)\) group elements
in eqn.(\ref{eq:3.10}) commute.

At this point it is worth mentioning the fact that whilst one cannot write down
the energy E appearing in the GCPF \(Z\) for arbitrary configurations of
\( i\) and \(j\) type vortices, it is possible for those configurations given
purely in terms of \( i\) or \( j\) types only. The resulting energy
\(E \) is just the generalization of eqn.(\ref{eq:3.13}), corresponding to
two Coulomb gas systems of \(i\) and \(j\) type vortices respectively.
Viewed in this way, the GCPF \(Z\) describes how these two different,
 `non-commuting' Coulomb gases
interact with each other. The relative strength of each Coulomb system is
controlled by the fugacities \({\sl z}_i \) and \( {\sl z}_j \).
One can imagine two extreme cases where \({\sl z}_i \gg {\sl z}_j \)
or \( {\sl z}_j \gg {\sl z}_i \), so that \(Z\) is dominated by \(i\) or
\(j\) types. Each of these situations  correspond to a kind
of abelian limit of \(Z\) where we would the usual
K-T picture of  a bound medium or a  plasma of vortices of
either type.

Since \({\sl z}_i \) and \({\sl z}_j \) are expressed in terms of the
core energies \({E}_i^c \) and \({ E}_j^c \) (see eqn.(\ref{eq:3.8.6}),
one has to check that these two abelian regimes can be actually realized.
Because we have used the kinematic arguments given earlier to focus our
attention on \(i\) and \(j\) types only, the vacuum parameter
\(a\) is implicitly taken to be in the range \( (-2, -1/2)\). It is then
clear from the \(a\) dependence of the core energies (eqn.(\ref{eq:3.6})),
that for sufficiently small temperature T,
the abelian limits
 \({\sl z}_i \gg {\sl z}_j \) and  \({\sl z}_j \gg {\sl z}_i \)
 correspond to the values \(a \rightarrow
-2 \) and \( a \rightarrow -1/2 \). For \(a \) = -1 in the mid range
of \((-2,-1/2) \), \({\sl z}_i \,=\, {\sl z}_j \) and we are in the
extreme nonabelian regime where both vortex types are equally likely
to be found in the system. Clearly this region is the most relevant to
investigating nonabelian effects, and clearly explains why we chose
\(a\) = -1 in the numerical plots given in Fig.2
and Fig.5.

Let us continue to study the 2-pair contributions to \(Z\) by
 considering  the case when the two vortex-antivortex pairs are
of different types i.e. one of type \( i \), the other \( j \).
We expect a very different answer in this case since now the
full nonabelian structure of the theory should be revealed
via the non-commutativity of the group elements \( {\cal G}_{i} ,
{\cal G}_{j} \). In the same manner as above, the form of
\( \Phi \) for a pair of type \( i \) at the points \( z_1 , z_2 \),
and pair of type \( j \) at \( z_3 , z_4  \), denoted
by \( {\Phi}_{i\,j}^{(1)} \) will be
\bear\label{eq:3.12}
\Phi_{i\,j}^{(1)} (z, z_1,\cdots,z_4)&=&
		\ \ \,  {\cal G}_{i}(z, z_1)\, {\cal G}_{i}^{-1}(z,z_2)\,
			{\cal G}_{j}(z, z_3) \,{\cal G}_{j}^{-1}(z, z_4)
			\nonum
	      &&        \qquad \times
			\Phi_{diag}(z, z_1, z_2 , z_3, z_4 )
			\nonum
	      &&        \times
			{\cal G}_{j}(z, z_4)\, {\cal G}_{j}^{-1}(z, z_3)\,
			{\cal G}_{i}(z, z_2)\,{\cal G}_{i}^{-1}(z, z_1) \nonum
\enar
In the GCPF we also must sum over other configurations related
to \(\Phi_{i\,j}^{(1)}\) by permutations of
the various group elements, keeping the ordering of the
points \( z_1 ,..., z_4 \) fixed. However,
since \( {\cal G}_{i} \) and \( {\cal G}_{j} \) do not
commute, the  energy of these different configurations are
not in general, simply related to each other. Also they are
much more complicated than the simple Coulomb
potential we saw previously, which is to be expected since the
latter is really a property of abelian vortex systems. Because
of this complexity it is not so straight forward to compute
explicitly the energy of each configuration. The difficulty evident
from the form of the energy density, which we will now discuss.

For the configuration given by \( \Phi_{i \, j }^{(1)} \)
of eqn.(\ref{eq:3.12}) the energy density, denoted by
\( {\cal E}_{i(-i)j(-j)}\) ,
is found to be
\bear\label{eq:3.13}
{\cal E}_{i(-i)j(-j)}
&=&     \biggl[ 2 (\varphi_1+ 2\varphi_2)^2
		- \,6  \varphi_1\,(2\,\varphi_1
		+\varphi_2) {\sin^2 [A(z- z_3)-A(z-z_4)]}
	\biggr ]
	\nonum
&&      \quad \times
		[ A_z(z-z_1)- A_z(z-z_2)]
		\,[ A_\zb(z-z_1)- A_\zb(z-z_2)]
	\nonum
&&              +\,2  (2\varphi_1+ \varphi_2)^2
	\nonum
&&      \quad \times
		\,[ A_z(z-z_3)- A_z(z-z_4) ]
		\,[   A_\zb(z-z_3)-  A_\zb(z-z_4)] \nonum
\enar
where in eqn.(\ref{eq:3.13}), \( A(z - z_s )
\, \equiv\,  Arg(z - z_s) \, , s = 1,...,4 \), and the subscript
\(z \,(\,\zb ) \) on \(A\) denotes differentiation by
\( \frac{\partial}{\partial z}\,(\frac{\partial}{\partial \zb})\)
respectively. In computing \( {\cal E}_{i(-i)j(-j)} \), we
have suppressed terms involving derivatives of \(\vf_1\) and \(\vf_2\)
as well as those involving the potential \(V\). They will only
contribute to the core energies \(  E^{c}_{\alpha} \) as we
discussed earlier (see eqn.(\ref{eq:2.16})).
It is interesting to observe in this equation, that
an effective interaction is induced between the
pair of type \( i \) and \( j \)  given by the \( {{\rm (sine)}^2} \)
term. As a check on the correctness of eqn.(\ref{eq:3.13}),
one may readily verify that \( {\cal E}_{i(-i)j(-j)} \) goes over
to the energy density of a single \(i\) (\( j\) ) type
pair if one takes the limit \(z_3 \rightarrow z_4 \)
(\( z_1 \rightarrow z_2 \) ) respectively.

Now in principle we would have to calculate the energy of another 255
possible orientations of the matrices \( {\cal G}_{i} ,
{\cal G}_{j} \) and their inverses which corresponds to all
possible permutations of the `charge' assignments
at the points \( z_1,...,z_4 \). However as we shall show below,
one can employ various similarity transformations acting on
the elements \( {\cal G}_{\alpha} \) to connect the energy of
one configuration to another. In this way we will see that one
need only calculate explicitly the energy density of two other orientations
other than the one described by \( \Phi_{i\,j}^{(1)}   \).

Denote by \( E[{\cal G}, <{\Phi}_{diag} >  ] \) the energy
of an arbitrary orientation of the four matrices
\( {\cal G}_{i}, {\cal G}_{-i}, {\cal G}_{j}, {\cal G}_{-j} \)
at the points \( z_1,...,z_4 \) . Then for any constant matrix
\( U \), we have
\beq\label{eq:3.14}
	E[ U {\cal G} {U}^{-1}, < {\Phi}_{diag}  >]
\, = \, E[ {\cal G} , U < {\Phi}_{diag} >{U}^{-1} ]
\eeq
Now consider the particular \( U \) matrices listed in Table 2.
They generate the transformations \( (i,j) \rightarrow (-i,-j),
(-i,j)\) and \( (j,i) \) on the \(i\) and \(j\) types of vortices, and hence
generate permutations of the \( {\cal G}_{\alpha} \)'s. Using these
transformations,
it turns out that we can relate the
energy of an arbitrary orientation of \( {\cal G}_{\alpha} \)'s
to those of just three particular orientations; \(\Phi_{i\,j}^{(1)}\)
(already given in eqn.(\ref{eq:3.12})) together with
\(\Phi_{i\,j}^{(2)},\,\Phi_{i\,j}^{(3)}\) defined by
\begin{eqnarray*}\label{eq:3.15}
\Phi_{i\,j}^{(2)} (z,z_1,\cdots,z_4)
	      &=& \ \ \,{\cal G}_j(z,z_1)\,{\cal G}_{-i}(z,z_2)\,
			{\cal G}_i(z,z_3)\,{\cal G}_{-j}(z,z_4)
			\nonum
	      &&        \qquad \times
			\ \Phi_{diag}(z,z_1,z_2,z_3,z_4)
			\nonum
	      &&        \times
			{\cal G}_j(z,z_4)\,{\cal G}_{-i}(z,z_3)\,
			{\cal G}_i(z,z_2)\,{\cal G}_{-j}(z,z_1)
\end{eqnarray*}
and
\bear
 \Phi_{i\,j}^{(3)} (z,z_1,\cdots,z_4)
	      &=& \ \ \,{\cal G}_i(z,z_1){\cal G}_{j}(z,z_2)
			{\cal G}_{-i}(z,z_3){\cal G}_{-j}(z,z_4)
			\nonum
	      &&        \qquad \times
			\ \Phi_{diag}(z,z_1,z_2,z_3,z_4)
			\nonum
	      &&        \times
			{\cal G}_j(z,z_4)\,{\cal G}_{i}(z,z_3) \,
			{\cal G}_{-j}(z,z_2)\,{\cal G}_{-i}(z,z_1)
	\nonum
\enar
The energy densities \( {\cal E}_{ij(-j)(-i)} \) and
\( {\cal E}_{ij(-i)(-j)} \) of these last  configurations,
(again in the region where \( |z - z_s | > {r_c} \) so that
one may approximate \( \Phi_{diag} \) by \( < \Phi_{diag} >\)),
are found to be
\bear\label{eq:3.16}
{\cal E}_{ij(-j)(-i)} \label{eq:3.18}
&=&
	\Bigl\{         2\,  ( 2\,\vf_1+\,\vf_2)^2
			+\,\sin^2[A(z-z_2)-A(z-z_3)]\,
	\nonum
&&      \times
			(\, 6\, (2\, \vf_1+\,\vf_2)
			\,{ \vf_2}\,\sin^2 [A(z- z_2)]
	\nonum
&&      \qquad
			-6\, \vf_1\, (\,\vf_1+2\,\vf_2)  )
	\Bigl\}         \,   A_z(z-z_1)  A_\zb(z-z_1)
	\nonum
&&      -
	\Bigl\{         4\, (2\,\vf_1+ \vf_2)^2
			\, \cos [A(z- z_2) - A(z- z_3 )]
	\Bigl\}
	\nonum
&&      \quad\times
			\,[  A_z(z - z_4)\,  A_\zb(z- z_1)  +
			  A_\zb(z - z_4)\,  A_z(z- z_1) ]
	\nonum
&&      +
	\Bigl\{         6\,\vf_2\, (2\,\vf_1
			+\, \vf_2) \,\sin [A(z-z_2) - A(z-z_3)]
			\, \sin[2 A( z-z_4)]\,
	\Bigl\}
	\nonum
&&      \quad\times
			\,[ A_z( z-z_1)   [A_\zb(z - z_2) -A_\zb(z-z_3)]
	\nonum
&&      \qquad\times
			A_\zb( z-z_1)   [A_z(z - z_2) -A_z(z-z_3)] ]
	\nonum
&&      +
	\Bigl\{         2\, ( \vf_1+ 2\,\vf_2)^2
			-6\,\vf_2\, (2\,\vf_1
			+\, \vf_2) \,\sin^2 [A( z -  z_4  )]
	\Bigl\}
	\nonum
&&      \quad\times
			[ A_z(z - z_2) -  A_z(z-z_3)]
			   [  A_\zb(z - z_2) -  A_\zb(z-z_3) ]
	\nonum
&&      +
			2\, (2\,\,\vf_1\,+\,\vf_2)^2
			\,\, A_z(z-z_4)  A_\zb(z - z_4 )
\enar
and
\bear
{\cal E}_{ij(-i)(-j)}  \label{eq:3.19}
&=&
%9
			2\,\Bigl\{ \,  (\vf_1 + 2\, \vf_2 )^2
			- 6\, \vf_2\, ( 2\, \vf_1 +  \vf_2 )  \,
			\sin^2 [A(z -  z_4)]
	\nonum
&&      \qquad
			- 6\,\sin^2 (z -  z_2)\,
	\Bigl[          \vf_2\, ( 2\, \vf_1 +  \vf_2 )  \,
			\cos [2\,A( z -  z_4 )]
			-\,\sin [A(z -  z_3)]^2\,
	\nonum
&&      \qquad\qquad\times
		\Bigl( \vf_1\, (  \vf_1 + 2\, \vf_2 )
			- \vf_2\, ( 2\, \vf_1 +  \vf_2 )  \,
			\sin^2 [A(z -  z_4)]
		\Bigl)
	\Bigl]
	\nonum
&&      \qquad
			+\, 3\, \vf_2\, ( 2\, \vf_1
			+  \vf_2 )  \,\cos [A(z -  z_3)]\,
			\sin [2\,A(z -  z_2 )] \,
	\nonum
&&      \qquad\times
			\sin [2\,A(z -  z_4 )]
	\Bigl\}         \, A_z(z-z_1)  A_\zb(z - z_1 )
	\nonum
&&                      + \,
%6
	\bigl\{         2  \left( 2  \vf_1 +  \vf_2 \right) ^2  -
			6 \sin^2 [A(z -  z_3) ]
		\bigl[
			\vf_1  (\vf_1
	\nonum
&&      \qquad
			+  2 \,\vf_2)  -  \vf_2
			 \left( 2  \vf_1 +  \vf_2 \right)
			\sin^2 [A(z -  z_4)]
		\bigl]
	\bigl\}\ \,
	\nonum
&&      \quad\times
			A_z(z -  z_2)  A_\zb(z- z_2)
	\nonum
&&      +\,
%3
	\bigl\{         2  \left(  \vf_1 + 2  \vf_2 \right) ^2
			- 6  \vf_2  \left( 2\vf_1 +  \vf_2 \right)
			\sin^2 [A(z -  z_4)]
	\bigl\}
	\nonum
&&      \quad\times
			A_z(z -  z_3)  A_\zb(z - z_3 )
	\nonum
&&      +\,
%2
	2\bigl\{         \left( 2  \vf_1
			+  \vf_2 \right) ^2  \bigl\}
			\,A_z(z -  z_4)  A_\zb(z - z_4 )
	\nonum
&&
%4
			-\,6\, \bigl\{ \vf_2  ( 2  \vf_1
			+  \vf_2 )   \sin [A(z -  z_3)]
			\sin [2 \,A( z -  z_4 )] \ \,
	   \bigl\}
	\nonum
&&      \quad\times
			\,[ A_z(z -  z_2)  A_\zb(z -  z_3) +
			  A_\zb(z -  z_2)  A_z(z -  z_3) ]
	\nonum
&&
%5
			-\,4\, \bigl\{  \left( 2  \vf_1
			+  \vf_2 \right) ^2  \cos [A(z -  z_3)]
	\ \,\bigl\}
	\nonum
&&      \quad\times
			\,[ A_z(z - z_2) A_\zb(z -  z_4) +
			  A_\zb(z - z_2)   A_z(z -  z_4)]
	\nonum
&&      +\,
%7
			6\,\Bigl\{\,\sin [A(z -  z_2)]
			\,\sin [2\,A( z -  z_3) ] \,
	\nonum
&&      \qquad\times
	\bigl[          - \vf_1\,( \vf_1 + 2\, \vf_2 )
			+ \vf_2\,( 2\, \vf_1 +  \vf_2 ] \,
			\sin^2[A (z -  z_4) ]
	\bigl]
	\nonum
&&      \qquad
			+\, \,\vf_2\,( 2\, \vf_1
			+  \vf_2 ) \,\cos [A(z -  z_2)]\,
			\sin[A (z -  z_3)]
	\Bigl\}
	\nonum
&&      \quad\times
			\,\sin [2\,A( z -  z_4 )]
			\,[  A_z(z-z_1)  A_\zb(z-z_2) +
			   A_\zb(z-z_1)    A_z(z-z_2) ]
	\nonum
&&
%8
	-4
	\,\Bigl\{       ( \vf_1 + 2\, \vf_2 ) ^2\,\cos [A(z -  z_2) ]+
			12\, \vf_2\,( 2\, \vf_1
			+  \vf_2 )
	\nonum
&&      \qquad \qquad
	\times          \,\sin [A(z -  z_4)]\,
	\Bigl[          - \cos [A(z -  z_3)]\,\cos [A(z -  z_4)]\,
	\nonum
&&      \qquad \qquad\times
			\sin [A(z -  z_2)  ]+\,
			\cos [A(z -  z_2)]\,\sin [A(z -  z_4)]
	\Bigl]
	\Bigl\}
	\nonum
&&      \qquad \qquad\times
			\,\,[  A_z(z-z_1)\, A_\zb(z-z_3) +
			     A_\zb(z-z_1)\,   A_z(z-z_3) ]
	\nonum
&&      +\,
%1
	4
	\bigl\{         \left( 2  \vf_1 +  \vf_2 \right) ^2
			\sin [A(z -  z_2)]
			\sin [A(z - z_3)]\,
	\bigl\}
	\nonum
&&      \quad\times
			\,[  A_z(z -  z_1)  A_\zb(z -  z_4) +
			   A_\zb(z -  z_1)    A_z(z -  z_4)   ]
\enar

Again it can be checked that the expression for \( {\cal E}_{i(j)(-j)(-i)}\)
in eqn.(\ref{eq:3.18}) goes to that of a single vortex-antivortex
pair of type \(i\) in
the limit \( z_2 \rightarrow z_3 \). However, there does not exist any
similar simple check on the expression  \( {\cal E}_{ij(-i)(-j)}\),
involving limits as some of the points \( z_1,..,z_4 \) are coincident.
This is because the orientation of \(SO(3)\) elements juxtaposes
non-commuting elements. For the same reasons, \( {\cal E}_{ij(-i)(-j)}\)
has the most complicated form of the three orientations
\(\Phi^{(1)}_{ij},...,\Phi^{(3)}_{ij} \).
In Fig.5a, the energy density \( {\cal E}_{i(-i)j(-j)} \)
has been plotted for \( \l =  2 , a = -1 \) where the  vortices are placed at
the points (-3,3), (3,3), (-3,-3), (3,-3) in the plane. For comparison,
Fig.5b shows a plot of
the energy density \( {\cal E}_{i(-i)i(-i)} \) for  abelian \(i\)-type
vortices, at the same positions, interacting Coulombically.
In both plots we have approximated \(\vf_1\) and \(\vf_2\) for
\(|z-z_2| < \,r_c \) by the trial functions given in eqns.(\ref{eq:2.15}).
It is clear that energy is localized at the centre of each vortex,
which is further evidence that core regions exist. Also apparent in the
nonabelian case is a suppression of the energy density of \( j\) type
with respect to \(i\) type vortices, which is a consequence of the
\({\rm (sine)}^2 \) terms in eqn.(\ref{eq:3.13}).

In the next section we shall outline a method for computing the
the energy of these 3 basic configurations, from the
corresponding energy density. As we shall see, this demonstrates that the
interaction energy between non-commuting vortices is quite different
and rather more complicated than the simple logarithmic Coulomb
potential that appears in the abelian case.

\sect{ Interaction potential between i and j-type \newline vortices }

In this section we shall present a method for calculating the potential
energy of the 4-vortex configurations, whose energy densities were given in
eqns.(\ref{eq:3.13}), (\ref{eq:3.18})
and (\ref{eq:3.19}) of the previous section. This
involves deriving  a set of differential equations in the variables
\( {z_1},...,{z_4} \) . For definiteness we shall consider the simplest
such configuration  (\ref{eq:3.12}),
with energy density \( {\cal E}_{i(-i)j(-j)} \) . The
total energy \mbox{\( {{ E} }_{i(-i)j(-j)} = \int {d}^2 {z}
\,{\cal E}_{i(-i)j(-j)}  \)} can be written as
\bear\label{eq:4.1}
{{E}}_{i(-i)j(-j)}(z_1 ,z_2,z_3,z_4 )& =&
{E}_{11} + E_{22} + E_{12}  + E_{21}
\nonum
&+& \pi\, <  (\varphi_1+ 2\varphi_2)^2 > \ln (\frac{|z_1- z_2 |}{r_c} )
       \,+\, 2\, E^c_i
\nonum
&+& \pi\,    < (2\varphi_1+ \varphi_2)^2 > \ln (\frac{|z_3 - z_4 |}{r_c} )
	\,+\, 2\, E^c_j
\enar
where
\bear\label{eq:4.2}
E_{11} \,& = & -  6\,< \varphi_1\,(2\,\varphi_1  +\varphi_2)>
 \int {d}^2 z \{ A_z(z-z_1) A_\zb(z - z_1 )
\nonum
&\times & {\rm sin}^2
 (A(z-z_3) - A(z-z_4)) \}
\nonum
\nonum
E_{22} \,&=&   -  6\,< \varphi_1\,(2\,\varphi_1
		+\varphi_2)>
\int {d}^2 z \{ A_z(z-z_2) A_\zb(z - z_2 )
\nonum
&\times& {\rm sin}^2
 (A(z-z_3) - A(z-z_4)) \}
\nonum
\nonum
E_{12} \,& = &
		-  6\,< \varphi_1\,(2\,\varphi_1
		+\varphi_2)>
\int {d}^2 z \{ A_z(z-z_1) A_\zb(z - z_2 )
\nonum
&\times& {\rm sin}^2
 (A(z-z_3) - A(z-z_4)) \}
\nonum
\nonum
E_{21} \, &=&
		-  6\,< \varphi_1\,(2\,\varphi_1
		+\varphi_2)>
\int {d}^2 z \{ A_z(z-z_2) A_\zb(z - z_1 )
\nonum
&\times& {\rm sin}^2
 (A(z-z_3) - A(z-z_4)) \}
\enar

The last four terms in eqn.({\ref{eq:4.1}) represent contributions from
isolated vortex-antivortex pairs of type \(i\) and \(j\) respectively.
This formula also takes into account contributions from within the
core regions, which gives rise to the core energies \( E^c_i\)
and \( E^c_j\) in this equation. \(E_{11},...,E_{21} \) on the other hand,
represent interactions amongst those pairs.

Using the relation \( \dztb (1/ (z_1 - z_2 ) ) =  \dztb \dzo A(z_1 - z_2 ) =
{\delta}^{(2)} ( z_1 -  z_2 ) \),
one can derive the following first order differential equations
for \( E_{12} \) and \( E_{21} \)
\bear\label{eq:4.3}
\dzt E_{12} \, &=&\,
		- \frac{ 6\,< \varphi_1\,(2\,\varphi_1
		+\varphi_2)>}{(z_1 - z_2 )} {\rm sin}^2 (A(z_2 -z_3 ) -
				      A(z_2 -z_4 ))
\nonum
\dzo E_{21} \, &=&\,
		- \frac{ 6\,< \varphi_1\,(2\,\varphi_1
		+\varphi_2)>}
{(z_2 - z_1 )} {\rm sin}^2 (A(z_1 -z_3 ) -  A(z_1 -z_4))
\enar
Once we have obtained solutions for \( E_{12} \) and \( E_{21} \), those
of \( E_{11}  \) and  \( E_{22} \) follow by taking the appropriate limit
\( z_1 \rightarrow z_2 \). Consider first the solution for \( E_{12} \),
which can be obtained by integrating  eqn.(\ref{eq:4.3}),
\bear\label{eq:4.4}
E_{12} (z_1,z_2,z_3, z_4 ) \,& =& \,
		-  6\, <  \varphi_1\,(2\,\varphi_1
		+\varphi_2) >
\nonum
&&\times\int\frac{{\rm d}z_2 }{(z_1 - z_2)}{\rm sin}^2 (A(z_2 - z_3 ) - A(z_2 -
z_4 ) )
			     + H_{12}
\enar
where the function \( H_{12} \) satisfies \( \dzt H_{12} = 0 \). Let \( I_{12}
\)
denote the integral in eqn.(\ref{eq:4.4}). It is straightforward to
write this integral in the following manner
\beq\label{eq:4.5}
    I_{12} \, = \, \int \frac{ {\rm d} \zp }{\acute{z} }\bigl\{ - \frac{1}{4}
       \frac{(\zp + \omo ) (\zpb +\omtb )}{(\zpb + \omob ) (\zp +
\omt )}  + \frac{1}{4} \frac{( \zpb + \omob ) (\zp + \omt )}{(\zp + \omo )
( \zpb + \omtb )} \bigr\}
\eeq
where  \( \omo \equiv z_1 -z_3 \) and \(\omt \equiv
z_1 -z_4 \).
Calculating the  integrals in eqn.(\ref{eq:4.5}), \( I_{12} \) is given by
\bear\label{eq:4.6}
I_{12} \, &=& \, \frac{1}{4} \ln (z_1 - z_2 ) \biggl\{ - 2 +
\frac{(\ztb - \zfb )(z_1 - z_3 )}{(\ztb - \zthb )(z_1 -z_4) } +
\frac{(\ztb - \zthb )(z_1 - z_4)}{(\ztb - \zfb )(z_1 - z_3 ) }\biggr\}
\nonum
&+&   \frac{(\ztb - \zthb )(z_4 - z_3 )}{4(\ztb - \zfb )
(z_1 - z_3 ) } \ln (z_2 - z_4 )+  \frac{(\ztb - \zthb )(z_4 - z_3 )}
{4(\ztb - \zfb )(z_1 - z_3 )} \ln (z_2 - z_4 )
\enar
In a  similar manner, we can solve for \( E_{21} \)
\beq\label{eq:4.7}
E_{21} \, = \,
		-  6\,< \varphi_1\,(2\,\varphi_1
		+\varphi_2)>
I_{21} (z_1, z_2, z_3, z_4 ) + H_{21}
\eeq
with \( \dzo H_{21} = 0 \). The integral \( I_{21} \) in
eqn.(\ref{eq:4.7}) is given by \( I _{12} \) of eqn.(\ref{eq:4.6} )
with \( z_1 \leftrightarrow z_2 \).

What remains is for us to determine the function \( H_{12} \). ( \( H_{21} \)
is again equal to \( H_{12} \) under the exchange \( z_1 \leftrightarrow
z_2 \) ). To determine \( H_{12} \) we differentiate \( E_{12} \)
once with \( \dzob \) , and obtain the differential equation
\beq\label{eq:4.8}
\dzob H_{12} \, = \,
	  \frac{      3\,< \varphi_1\,(2\,\varphi_1
		+\varphi_2)>}{ (\zob - \ztb ) }\bigl\{ 2 -
\frac{(z_1 - z_3 )(\zob - \zfb )}{(\zob - \zthb )(z_1 - z_4 )}
- \frac{(\zob - \zthb )(z_1 - z_4 )}{(z_1 - z_3 )(\zob - \zfb )} \bigr\}
\eeq
Solving for \( H_{12} \) one finds
\bear\label{eq:4.9}
H_{12} \,& = &\,
		3\,< \varphi_1\,(2\,\varphi_1
		+\varphi_2)>
 \biggl\{ \frac{(z_1 - z_4 ) (\zthb - \zfb )}{(
z_1 -z_3 )(\ztb - \zfb )} \ln (\zob - \zfb )
	\nonum
	&+& \frac{(z_1 - z_3 )(\zfb - \zthb )}
		{(z_1 - z_4)(\ztb - \zthb )} \ln ( \zob - \zthb )
		-\frac{1}{2} \ln (\zob - \ztb )
	\nonum
	&+&
		 \frac{(\ztb - \zfb )(z_1 - z_3 )}{
		(\ztb - \zthb )(z_1 - z_4 ) } \ln (\zob - \ztb )
		\frac{(\ztb - \zthb )(z_1 - z_3 )}
		{(\ztb - \zfb )(z_1 - z_3 )} \ln (\zob - \ztb ) \biggr\}
\enar
Defining  \( {{\rm E} }^{int} = E_{12} + E_{21} \)  as the interaction
energy between the vortex-antivortex pair of type \( i \) at \( (z_1 , z_2 ) \)
in the presence of the pair of type \( j \) at \( (z_3,z_4 ) \), one
finds
\newpage
\bear\label{eq:4.10}
{{\rm E}}^{int} (z_1, z_2, z_3, z_4 ) \,
	& = &   \, -\frac{3}{2} \,< \varphi_1\,(2\,\varphi_1   +\varphi_2)>
	\biggl\{        -4 \ln (\frac{| z_1 - z_2 |}{R})
	\nonum
	& +&
		 \frac{(\ztb - \zfb )}{ (\ztb - \zthb ) } \bigl\{ \ln (\frac{ |z_2 - z_4
|}{R} )
		+\frac{(z_1 - z_3 )}{(z_1 - z_4 ) } [\ln (\frac{|z_1 - z_2 |}{R} )
		- \ln (\frac{|z_2 - z_4 |}{R} ) ]  \bigr\}
	\nonum
	&+&
		 \frac{(z_2 -z_4 )}{(z_2 - z_3 )} \bigl\{ \ln (\frac{|\ztb - \zfb |}{R} )
		+\frac{\zob - \zthb }{\zob - \zfb } [ \ln (\frac{|\zob - \ztb |}{R} ) -
		\ln (\frac{|\ztb - \zfb |}{R} ) ] \bigr\}
	\nonum
	& + &
		 \frac{\ztb - \zthb }{\ztb - \zfb } \bigl\{ \ln (\frac{|z_2 - z_3 |}{R} )
		+\frac{z_1 - z_4 }{z_1 - z_3 } [ \ln (\frac{|z_1 - z_2 |}{R} )
		- \ln ( \frac{|z_2 - z_3 |}{R} ) ] \bigr\}
	\nonum
	& +&
		 \frac{z_2 - z_3 }{z_2 - z_4 } \bigl\{ \ln (\frac{|\ztb -\zthb |}{R} )
		+\frac{\zob - \zfb }{\zob - \zthb } [ \ln ( \frac{|\zob - \ztb |}{R})-
		\ln ( \frac{|\ztb - \zthb |}{R} ) ] \bigr\}  \,
	\nonum
	& +& z_1 \leftrightarrow z_2
		\biggr\}
\enar

In eqn.(\ref{eq:4.10} ), we have introduced   an  infrared cutoff \( R\),
which
can be taken as the macroscopic size of the 2-dimensional system.
It will be an important check on the total configurational energy
\( {{\rm E}}_{i(-i)j(-j)} \), that one can take the thermodynamic limit \( R
\rightarrow \infty \) and still obtain finite total
energy. To verify this, the various logarithmic divergences in the
interaction energies given in eqn.(\ref{eq:4.10} ) have to cancel with
the self energies \( E_{11} \) and \( E_{22} \).
The latter are obtained from \( E_{12} \) and \( E_{21} \) in the limit
\(z_1 \rightarrow z_2 \) and \( z_2 \rightarrow z_1 \) respectively. Of
course in taking this limit one should remember that there is an
effective short-distance cutoff in the value
of \( |z_1 - z_2 | \) which is the core size \(r_c \). For
separations smaller than this, the fields \( \varphi_1 , \varphi_2 \)
are no longer given by their vacuum expectation values, but vanish
as \( |z_1 - z_2 | \rightarrow 0 \) to maintain the
finiteness of the energy, as discussed in Section 2.

With this in mind, it is straightforward to compute the
following terms proportional to
\( \ln \,  R \)  in \( E_{11} + E_{22} \),
\bear\label{eq:4.11}
 \frac{3}{2}\,\ln \,R  \,& < & \varphi_1\,(2\,\varphi_1
		+\varphi_2)>   \bigl\{  4 + \frac{\zob - \zfb }{\zob - \zthb }
		+ \frac{z_1 - z_4}{z_1 - z_3 }
		+ \frac{\zob -\zthb }{\zob - \zfb }
\nonum
&+& \frac{z_1 -z_3}{z_1 - z_4 } + \frac{\ztb - \zfb }{\ztb - \zthb  } +
\frac{z_2 - z_4 }{z_2 - z_3 }
+ \frac{\ztb - \zthb }{\ztb - \zfb} + \frac{z_2 - z_3}{z_2 - z_4 } \bigr\}
\enar

By comparing eqn.(\ref{eq:4.11}) with eqn.(\ref{eq:4.10}), one sees that indeed
the
\( \ln\,R \)  terms cancel, so that one may safely take the thermodynamic
limit. This result is the nonabelian generalization
of the well known property found in abelian vortex systems,
that only overall neutral configurations, with equal numbers of vortices and
antivortices  are relevant
 (i.e. have finite
energy) in the thermodynamic limit. Neutrality in the present context simply
means with respect to the group \( \IQ_8 \), so
that for every vortex of one type, we must include the antivortex of the same
type. This explains the restrictions on the allowed
total vorticities in the GCPF of eqn.(\ref{eq:3.9}) in the last section.

It is interesting to see in the formula for the interaction energy
eqn.(\ref{eq:4.10}) the usual Coulomb terms as well as new logarithm-like
terms, which differ from the former by the prefactors that are rational
functions of the points \(z_1,...,z_4 \).
One can, by similar methods to those described above, obtain
explicit expressions for the remaining configurations. Whilst
the result for \( E_{j(-j)i(-i)} \) is very similar to
that of \( E_{i(-i)j(-j)} \) described above, those of
\( E_{j(-i)i(-j)} \) and \(E_{ij(-i)(-j)} \) are particularly
complicated and lengthy, although they still involve the logarithm-like
functions seen above. Details of these expressions will be given
elsewhere [12].

\vspace{2ex}
\sect{Summary and Conclusion}
\vspace{2ex}
In this paper we have made some first steps in the investigation
of the statistical mechanical properties of nonabelian vortices
in two spatial dimensions. We have done this with particular reference
to vortices characterized by the fundamental group \( \IQ_8 \). An
ansatz was presented that describes isolated vortices, which as we showed in
the
appendix, does indeed describe configurations of minimum energy.
To eventually describe statistical mechanical properties, it is
necessary to discuss multi-vortex configurations, which would
be present in a realistic system such as nematic liquid crystals
[9], at finite temperature. We computed the energy density of
all relevant configurations with a maximum of four vortices,
and gave a method for calculating the total energy from this.

The kinematic arguments discussed in Section three simplified the
discussion somewhat, since they implied that only
two out of possibly four species of nonabelian vortices would be
relevant to the statistical mechanics of the nonabelian gas.
Even so, the form of the energy and energy density of
the many different 4-vortex configurations
are very much more complicated than the corresponding expressions
for abelian vortices . An important
check on these results was that the energy of these configurations remained
finite in the thermodynamic limit, in which the
size of the system becomes infinite.

As stated in the introduction, the results presented in this paper should
provide a starting point in which to investigate amongst other things,
nonabelian screening mechanisms which generalize those already investigated
for abelian vortices [1,6]. A step in this direction would be to derive
a nonabelian Poisson-Boltzmann like equation for the linearly screened
potential
between two test vortices, using the form of either the energy density
or energy as
given in  Section 3 and Section 4, [12].
There are many additional problems that one could
investigate in the context of nonabelian vortices and K-T like phase
transitions,
which have already been studied in the abelian case.
For example one could consider the effects of putting the system on a surface
of non-trivial topology e.g. a sphere [16]. In addition it would be interesting
to consider further modifications of the nonabelian system by including
 vortex induced Berrys phases. In the abelian case, the
presence of such phases was shown to have a dramatic effect   on the nature of
the
K-T phase transition [17] .

\newpage
\appendix
\noindent{\Large\bf Appendix }
\sect{}
\sect* {Stability of nonabelian vortices}
In Section 2 (see eqn.(\ref{eq:2.10})) and Table 1,
 we presented an ansatz for describing  isolated nonabelian
vortices described by the fundamental group \(\IQ_8 \).
Specifically these were of the form \( {\cal G }_{\alpha} (\theta )
{\Phi }_{diag}( r )  {\cal G }_{\alpha}^{-1} (\theta ) \), where
\( {\cal G}_{\alpha} (\theta ) \,, \alpha = i, j , k \) were rotations
about the \( x, y\) or \( z \) axes by angles \( \theta /2 \), (with the angles
\( \theta / 2 \) replaced by \(- \theta / 2 \)  for the
corresponding antivortices ). In this appendix we will give a proof that
these ansatzes are the physically correct ones in that they describe
vortices with lowest energy\footnote{We will do this explicitly for
 \(k\)-type vortices; the calculation for \(i\) and \(j\) types
follow in a very similar manner with the same conclusions concerning
stability. We will therefore only present the calculation for  \(k\)-type
vortices only.} To accomplish this we shall simply
deform our original ansatz by boosting the specific choice of
\( SO(3) \) element described above, by an arbitrary rotation
depending on three angles \( \alpha , \beta , \gamma \) which will be
functions of \( \theta \) and \( r \) in general,
and argue that such deformations always lead to
an increase in the vortex energy. The only constraint
that the boosts have to satisfy is that they preserve the boundary
conditions satisfied by the group elements \( {\cal G}_{\alpha} \)
as discussed in Section 2 of the paper (see eqn.(\ref{eq:2.11}) )
This simply guarantees that the closed curves in \( M \) which
\({\cal G}_{\alpha} \) parameterize, are homotopically invariant
under the deformations.
For example, a generalized ansatz for  vortices of type \(k \) will be
\beq\label{eq:A1}
	\Phi_{ijk}  =   {{\cal G}_i(\alpha) {\cal G}_j(\beta)
			{\cal G}_k(\gamma)\, \Phi_{diag}\,
			{\cal G}_k^{-1}(\alpha) {\cal G}_j^{-1}(\beta)
			{\cal G}_i^{-1}(\gamma) }
\eeq
where in eqn.({\ref{eq:A1}) \( \a,\,\b \)  are the single valued function of
\(\theta\), while\newline \mbox{\( \g (\theta = 0 ) = \g (\theta = 2 \pi ) +
\pi  \)}
, in order to preserve the boundary conditions as
discussed above.
In showing that perturbations of our original ansatz of eqn.(\ref{eq:2.10})
always increase the energy of a vortex, we shall restrict ourselves
to regions outside of the vortex core \( r_c \),
where the fields \(\varphi_1 (r) \) and \(\varphi_2 (r) \) can be approximated
by
their vacuum expectation values. Although the particular
numerical solutions described in Section 2 used our original
(unperturbed) ansatz for \({\cal G}_{\alpha}(\theta) \) , the
property that they asymptotically go to their vacuum values
should be true even for a modified ansatz, (see the discussion
after eqn.(\ref{eq:2.15}) of Section 2). Outside vortex cores,
the energy density \({\cal E}_{ijk}\) of the  field configuration
\( \Phi_{ijk} \) taken the form
\newpage
\bear\label{eq:A2}
	{\cal E}_{ijk}  & = &
			{ <\varphi_1}\,{ \varphi_2}>\,\left( 2\,{{\alpha '}^2}
			+ 6\,\cos (2\,\beta )\,{{\alpha '}^2}
			+ 8\,{{\beta '}^2}
			- 8\,\sin (\beta )\,\alpha '\,\gamma '
			- 4\,{{\gamma '}^2} \right)  \nonum +
		      &&
			{{<{ \varphi_2}}^2}>\,
		       \Biggl[
			{{7\,{{\alpha '}^2}}\over 2}
			+ {{3\,\cos (2\,\beta )\,{{\alpha '}^2}}\over 2}
			+ {{3\,\cos (2\,\left( \beta - \gamma  \right) )
				\,{{\alpha '}^2}}\over 4}
			+ {{3\,\cos (2\,\gamma )\,{{\alpha '}^2}}\over 2}
			\nonum
		      && \qquad
			+ {{3\,\cos (2\,\left( \beta  + \gamma  \right) )
				\,{{\alpha '}^2}}\over 4}
			- 3 \,\sin (\beta  - 2\,\gamma )\,\alpha '\,\beta '
			+ 3\,\sin (\beta  + 2\,\gamma )\,\alpha '\,\beta '
			\nonum
		      && \qquad
			+ 5\,{{\beta '}^2}
			- 3\,\cos (2\,\gamma )\,{{\beta '}^2}
			+ 4\,\sin (\beta )\,\alpha '\,\gamma '
			+ 2\,{{\gamma '}^2}
		      \Biggr]
		      \nonum +
		      &&
			{{<{ \varphi_1}}^2}>\,
		      \Biggl[
			{{7\,{{\alpha '}^2}}\over 2}
			+ {{3\,\cos (2\,\beta )\,{{\alpha '}^2}}\over 2}
			- {{3\,\cos (2\,\left( \beta  - \gamma  \right) )
				 \,{{\alpha '}^2}}\over 4}
			- {{3\,\cos (2\,\gamma )\,{{\alpha '}^2}}\over 2}
			\nonum
		      && \qquad
			- {{3\,\cos (2\,\left( \beta  + \gamma  \right) )
				\,{{\alpha '}^2}}\over 4}
			+ 3\,\sin (\beta  - 2\,\gamma )\,\alpha '\,\beta '
			- 3\,\sin (\beta  + 2\,\gamma )\,\alpha '\,\beta '
			\nonum
		      && \qquad
			+ 5\,{{\beta '}^2}
			+ 3\,\cos (2\,\gamma )\,{{\beta '}^2}
			+ 4\,\sin (\beta )\,\alpha '\,\gamma '
			+ 2\,{{\gamma '}^2}
		      \Biggr]
		      \nonum
   \enar
where in eqn.(\ref{eq:A2}) `\(\prime\)'indicates differentiation
with respect to coordinate \(\theta\) and the vacuum parameter \( a\).
Using the relation between \(< \vf_1> \) and \( <\vf_2> \)
\beq \label{eq: A4}
	<\vf_2>\,=\ a\ <\vf_1>
\eeq
we rewrite the energy density of eqn.(\ref{eq:A2}) as
\beq\label{eq:A5}
  {\cal E}_{ijk} \,=\, {\xi^i}' \ M_{ij} \ {\xi^j}' <\vf_1^2 >
\eeq
where
\beq\label{eq:A6}
 M_{ij}=\pmatrix{   h_1  &  q_1  &  r_1  \cr
		    q_1  &  g_1  &  0    \cr
		    r_1  &  0    &  p_1  \cr}
\eeq
\( {\xi^i}'= \a'\,,\,\b'\,,\gamma' \)   and
\bear\label{eq:A7}
h_1 &=&
	{1\ovr 4} [ ( 14 + 8\,a + 14\,{a^2} + 6\,\cos (2\,\beta ) +
	24\,a\,\cos (2\,\beta ) + 6\,{a^2}\,\cos (2\,\beta ) \nonum
     && - \quad
	3\,\cos (2\,\left( \beta  - \gamma  \right) ) +
	3\,{a^2}\,\cos (2\,\left( \beta  - \gamma  \right) ) -
	6\,\cos (2\,\gamma ) + 6\,{a^2}\,\cos (2\,\gamma ) \nonum
     && - \quad
	3\,\cos (2\,\left( \beta  + \gamma  \right) ) +
	3\,{a^2}\,\cos (2\,\left( \beta  + \gamma  \right) ) ] \nonum
g_1 &=&
	[ 5 + 8\,a + 5\,{a^2} + 3\,\cos (2\,\gamma ) -
	3\,{a^2}\,\cos (2\,\gamma ) ] \nonum
p_1 &=&
       2\,{{\left(  a -1 \right) }^2} \nonum
q_1 &=&
       3\,\left( {a^2}-1 \right) \,\cos (\beta )\,
       \sin (2\,\gamma ) \nonum
r_1 &=&
       2\,\left(  a -1 \right)^2\,\sin (\beta ).
\enar
Now, if one can show that all of the energy eigenvalues of this matrix
are positive, it would be sufficient to show the stability of our
original ansatz.
The characteristic polynomial \( C(\chi , \alpha,\beta,
\gamma, a ) \)
associated with \(M_{ij} \)is given by
\bear\label{eq:A8}
    {{\chi }^3} -
    {{\chi }^2}\,\left( { g_1} + { h_1} + { p_1} \right)  -
    \chi \,[ - { g_1}\,{ h_1}
    - { g_1}\,{ p_1} - { h_1}\,{ p_1} + {{{ q_1}}^2}
    + {{{ r_1}}^2} ] \nonum
    \quad
    + { p_1}\,{{{ q_1}}^2} + { g_1}\,{{{ r_1}}^2}
    - { g_1}\,{ h_1}\,{ p_1} & \ =\ C(\chi)\nonum
\enar
To analyze the possible signs of the eigenvalues \( \chi \)
 (which are solutions of the equation \( C(\chi ) = 0 \) ),
we shall first study the extremal points of \( C(\chi) \) with respect
to the variable
\( \chi \). The general form of \( C(\chi) \) is illustrated
in Fig.4. If we can show that the values of \(\chi \)
at the two extremal points ( denoted by \( {\chi}_{e1} ,
{\chi}_{e2} \) in Fig.4 ) are bounded below by zero, then it follows
that at least two of the three eigenvalues are also similarly bounded.
To show that the remaining eigenvalue is \( \geq 0 \), we need
only check the value of \( C \) at \( \chi = 0 \), is \( \leq 0 \).

The values of \( \chi \) at extremal points of \(C\) are
given by
\beq\label{eq: A9}
   3\,{{\chi }^2} - 2\,\chi \,
   [ { g_1} + { h_1} + { p_1} ] + ( { g_1}\,{ p_1}
   + { h_1}\,{ p_1}  +{ g_1}\,{ h_1}
   - {{{ q_1}}^2} - {{{ r_1}}^2} ) \ =\ 0
\eeq
This  has two roots given by
\beq\label{eq:A10}
     \chi = {1\ovr 3} \bigr [  (g_1 + h_1 + p_1)
	       \+-\sqrt{  (g_1 + h_1 + p_1)^2
	       - 3 ( g_1h_1+ g_1 p_1 +h_1p_1 - q_1^2 - r_1^2)}
	       \bigr ]
\eeq

To get real and positive solutions for \( \chi\) in (\ref{eq:A10})
, requires the following two
constraints
\bear\label{eq:A11}
 g_1 +  p_1 +h_1 &>& \ 0 \\
 \label{eq:A12}
 g_1h_1+ g_1 p_1 +h_1p_1 - q_1^2 - r_1^2 &>&\ 0
\enar

As we shall see below, constraint (\ref{eq:A11} ) is satisfied because each
of the functions \( p_1 , h_1 \) and \( g_1 \) are separately positive.
{}From the definitions of eqn.(\ref{eq:A7} ) \(p_1\) is manifestly positive,
so we will study \( g_1 \) and \( h_1 \) instead. Consider first \( g_1 \).
To find a lower bound on this function for arbitrary angles \(\alpha ,
\beta, \gamma \), we consider varying the vacuum parameter \(a \) to find out
if
\(g_1 \) has a minimum value. We find
\bear\label{eq:A13}
{{\partial g_1}\over{\partial a}}
&=&
	8 + 10\,a - 6\,a\,\cos (2\,\gamma )\\
a_{ext}
&=&
	{{-8}\over {10 - 6\,\cos (2\,\gamma )}}\\
{{\partial^2 g_1}\over{\partial a^2}}
&=&
	10 - 6\,\cos (2\,\gamma )
\enar
Obviously,
\({{\partial^2 g_1}\over{\partial a^2}}\ >\ 0 \)
so the extrema of \( g_1 \) found by varying the parameter \( a \) is
a minimum and
\beq\label{eq:A14}
 g_1\mid_{min} =
	{{9\,{{\sin (2\,\gamma )}^2}}\over {5 - 3\,\cos (2\,\gamma )}}.
\eeq
Hence it is clear that \( g_1 \geq 0 \)

Moving on to the function \( h_1 \), we can apply the same ideas and
find that
\bear\label{eq:A15}
{{\partial h_1}\over{\partial a}}
&=&
	2 + 7\,a + 6\,\cos (2\,\beta ) + 3\,a\,\cos (2\,\beta ) +
	{{3\over 2}\,a\,\cos (2\,\left( \beta  - \gamma  \right) )}
\nonum
& +& 3\,a\,\cos (2\,\gamma )  \quad + {{3\over 2}\,a\,
	\cos (2\,\left( \beta  + \gamma  \right) )}\\
a_{ext}
&=&
	{{-4 - 12\,\cos (2\,\beta )}\over
	{14 + 6\,\cos (2\,\beta ) + 3\,
	\cos (2\,\left( \beta  - \gamma  \right) ) + 6\,\cos (2\,\gamma ) +
	3\,\cos (2\,\left( \beta  + \gamma  \right) )}}
\nonum
{{\partial^2 h_1}\over{\partial a^2}}
&=&
	7 + 3\,\cos (2\,\beta ) + {{3\,
	\cos (2\,\left( \beta  - \gamma  \right) )}\over 2} +
	3\,\cos (2\,\gamma ) + {{3\,\cos (2\,\left( \beta  +
	\gamma  \right) )}\over 2}
\nonum
\enar
It is easy to see that
\({{\partial^2 h_1}\over{\partial a^2}}\ >\ 0 \)
so again \( a_{ext} \)  describes  a minimum of \( h_1 \) with
\beq\label {eq:A16}
h_1\mid_{min}=
	{{9\,{{\cos (\beta )}^2}\,\left( 18 - 14\,\cos (2\,\beta ) -
	\cos (2\,\beta  - 4\,\gamma ) - 2\,\cos (4\,\gamma ) -
	\cos (2\,\beta  + 4\,\gamma ) \right) }\over
	{2\,\left( 14 + 6\,\cos (2\,\beta ) +
	3\,\cos (2\,\left( \beta  - \gamma  \right) ) +
	6\,\cos (2\,\gamma ) + 3\,\cos (2\,\left( \beta  + \gamma  \right) )
	\right) }}
\eeq

By inspecting the numerator in eqn.(\ref{eq:A16}), one can convince oneself
that again \(h_1\mid_{min} \geq 0 \). Thus we have shown that constraint
(\ref{eq:A11}) is satisfied.

We use the same ideas as above, to prove that eqn.(\ref{eq:A12}) is always
satisfied. It will be sufficient to prove the bound
\beq\label{eq:A17}
 h_1p_1  - q_1^2 - r_1^2 \ > \ 0
\eeq
since we have already shown the positivity of \(h_1,g_1 \) and \( p_1\).
In fact, after some manipulations one can simplify the
expression for
\(h_1  p_1 - q_1^2 -r_1^2 \)   and obtain
\beq\label{eq:A18}
h_1p_1-q_1^2-r_1^2
=
	      {2(a-1)^2}\biggl( {{{{\cos^2 (\beta )}}\ovr 2}
	\,{{( -1 + a + 3\,\cos (2\,\gamma ) +
	 3\,a\,\cos (2\,\gamma ) ) }^2}}\biggr)
\eeq
which is now in manifestly positive form. Hence the second constraint
eqn.(\ref{eq:A12}) is also satisfied, and so, as explained earlier, we are
guaranteed that at least two of the three eigenvalues \( \chi \)
are positive. To prove that the third eigenvalue is itself non-negative,
it is sufficient to check the value of the characteristic polynomial
\( C(\chi ) \) at \( \chi = 0 \). We find
\bear\label{eq:A19}
C(\chi)\mid_{\chi=0}
&=&
	-(  g_1\, h_1\, p_1 )
	+  p_1\,{ q_1}^2
	+ g_1\,{ r_1}^2
\nonum
&=&
	-8(a-1)^2\,(\,a\,+\,2\,)^2\,(\,2a\,+\,1\,)^2\cos^2(\beta)
\enar
which is a negative semi-definite quantity, so that we have the situation as
depicted in Fig.4 where all three eigenvalues are positive.
This completes the
proof that all perturbations of the original ansatz describing
isolated nonabelian \(k\)-type vortices increase the energy.
The vortices described by this ansatz are therefore stable. As
stated earlier, a similar analysis applied to \(i\) and \(j\) type vortices
shows that the corresponding ansatz also describe stable defects.
\newpage
\vspace{4ex}
\noindent{\large Acknowledgements}
\vspace{2ex}

 We thank K. Rama for collaboration during the initial stages of this work,
and D. Johnston for useful discussions.
We also thank H.-K. Lo for comments on the manuscript.
S.T. would like to thank the
Royal Society of  Great Britain for financial support.

\vspace{2ex}
\newpage  \noindent{\large References }
\vspace{2ex}
\begin{description}
\item{[1]} P. Minnhagen, {\it Rev. Mod. Phys.} {\bf 59} (1987) 1001.
\item{[2]} A. Vilenkin , {\it Phys. Rep.} {\bf 121} (1985) 263.
\item{[3]} L. Onsager, {\it Nvovo Cimeto } {\bf 6 }, {\it Supp.} {\bf 2}
		(1949) 249; R.P. Feynmann, {\it Progress in low temperrature
		physics}, {\bf Vol.1} ed. C.J. Gorter (North-Holland,
		Amsterdam, 1955) p.17; A.L. Felter,
		{\it Phys. Rev. Lett.} {\bf 27} (1971) 986,
		{\it Ann. Phys.} (N.Y.) {\bf 70} (1972) 67.
\item{[4]} L.P. Kadanoff, {\it J. Phys} {\bf A11} (1978), 1399.
\item{[5]} V.L. Berezinsky, {\it Zh. Eskp.
	Teor. Fiz. } {\bf 61} 1144  [{\it Sov. Phys. JETP } {\bf 34 }(1972)
	610.]
\item{[6]} J.M. Kosterlitz and D.J. Thouless, {\it J. Phys.
	}{\bf C6 }(1973)1181.
\item{[7]} D.R. Nelson and J.M. Kosterlitz,
	{\it Phys. Rev. Lett. }{\bf 39 }(1977) 1201.
\item{[8]} D.C. McQueeny, G. Agnolet and J.D. Preppy,
	{\it  Phys. Rev.Lett. }{\bf 52 }(1984) 1325.
\item{[9]} G.E. Volovik and V.P. Mineev, {\it Zh. Eskp.
	Teor. Fiz. } {\bf 72} 2256  [{\it Sov. Phys. JETP } {\bf 45 }(1977)
	1186.]
\item{[10]} J. Frolich in {\it Renormalization
	Theory}, Proc. of the NATO Advanced Study Institute, Erice 1975,
	eds. G. Velo and A.S.Wightman (Reidel, Dordrecht/Boston 1976,
	p.371); A.M. Polyakov, {\it Nucl. Phys.} {\bf 120} (1977)
	429; S. Samuel,{\it Phys. Rev. } {\bf D18} (1978)
	1916.
\item{[11]} P. Minnhagen, {\it Phys. Rev. }{\bf B23} (1981)
	5745.
\item{[12]} C. Kobdaj and S. Thomas, work in progress.
\item{[13]} H.-K. Lo and J. Preskill, "Nonabelian vortices and
	    nonabelian statistics", CALT-68-1867, hep-th 9306006.
\item{[14]} T.W.B. Kibble, {\it Phys. Rep.} {\bf 67}  (1980) 183;
	    J. Preskill and L. Krauss, {\it Nuclear Physics}
	    {\bf B341} (1990) 50.
\item{[15]} See chapter 4 of "Gauge Fields and Strings" by A.M. Polyakov,
	    Harwood Academic Publishers, 1987.
\item{[16]} B.A. Ovrut and S. Thomas, {\it Mod. Phys. Lett.}
	{\bf  A5} (1990) 2351; {\it Phys. Rev. } {\bf D 43} (1990) 1314.
\item{[17]} S. Thomas, {\it Nucl. Phys.} {\bf B386}  (1992) 592;
	{\bf B392} (1993) 619.
\end{description}

%----------------------------Caption----------------------------
\clearpage
{\large \bf { Table Captions}}
\begin{description}
\item{{\bf Table 1 :}}\\
	Table 1 lists the \(SU(2)\) group elements \(g\) of isolated
		 nonabelian vortices corresponding to elements \(h\),
		 of the discrete
		 fundamental group \( \IQ_8 \), which are also given. In
		 addition, the associated \( SO(3) \) group elements \({\cal G}\)
		 are listed.
\item{{\bf Table 2 :}} \\
	Table 2 shows how the similarity transformations generated
		 by the matrices \( U\):
		\({\tilde{{\cal G}_i}}\,=\, U\,{\cal G}_i\,U^{-1}\)
		 , \({\tilde{{\cal G}_j}}\,=\, U\,{\cal G}_j\,U^{-1}\)
		 , act on the \(SO(3) \) group elements of \(i\) and \(j\)
		 type vortices. Also given is the action of \(U\) on the fields
		 \( \Phi_{diag} \).

\end{description}
\newpage
{\large \bf { Figure Captions}}
\begin{description}
\item{{\bf Figure 1 :} }\\
	Figure 1 shows the ellipse described by the equation
		\mbox{\(\vf_1^2+\vf_2+\vf_1\,\vf_2\,=\,
		\frac{1}{(\lambda+\lambda')}\)} for various values of
		\(\lambda, \lambda'\), that corresponds to the vacuum
		expectation values \( < \Phi_{diag}>\)
		of the model considered in the text, when \(\rho\, =\,0 .\)
		When \(\rho\not=\,0, \ <\Phi_{diag}>\) takes on discrete
		values only, given by the intersection of the ellipses
		with the dashed radial lines in Fig.1. At these values
		of \(\Phi_{diag} \) as well as those given by the intersection
		of the dotted radial lines with the ellipses, there is an
		unbroken \( U(1) \) symmetry.
\item{{\bf Figure 2 :} }\\
	Figure 2 shows numerical solutions (dotted curves) and
		trial solutions (solid curves) for the fields \(\vf_1\)
		and \( \vf_2\) which minimize the energy of an isolated
		vortex of type \( i\), for \(\lambda\,=\,2,\,0.01;\,
		\lambda'\,=\,0\) and \( a\,=\,-1 \). Also indicated
		is the approximate position of the core region
		\(r\,\leq\,r_c\).
\item{{\bf Figure 3 :} }\\
	Figure 3 illustrates the relative positions of the regions
		\({\bf I}_{ij},\ {\bf I}_{jk}\) and \({\bf I}_{ki}\) of
		stable \(ij,\ jk\) and \(ki\) type vortices, with respect
		to the allowed values of \(<\Phi_{diag}>\) in the case
		\(\rho\,=\,0\).Boundary values of the vacuum parameters
		\(a\,\equiv\,<\vf_2>/<\vf_1>\) are also given.
\item{{\bf Figure 4:} }\\
	Figure 4 illustrates the form of the characteristics polynomial
		\( C(\chi)\) defined in appendix A, as a function of
		\( \chi\). \(\chi_0,\, \chi_1 \) and \(\chi_2\)
		(crosses in Fig.4) are the eigenvalues
		corresponding to the solutions of the equation
		\( C(\chi)\,=\,0\). \(\chi_{e1},\ \chi_{e2}\)
		denote the values of \(\chi\) at the turning points of
		\( C(\chi)\).
\item{{\bf Figure 5:} }\\
	Figure 5a is a plot of the energy density
		\({\cal E}_{i\,(-i)\,j\,(-j)}\) of four non abelian vortices
		at positions (-3,3), (3,3), (-3,-3) and (3,-3) in the
		\(x-y\) plane, with \(\lambda\,=\,2\), \(\lambda'\,=\,0\)
		and \(a\) = -1. For comparison, Fig.5b illustrates a
		plot of the energy density  \({\cal E}_{i\,(-i)i(-i)}\)
		corresponding to abelian vortices placed at the same
		position.
\end{description}

%----------------------------Table --------------------------------------
\renewcommand{\baselinestretch}{1}
\begin{table}
{\footnotesize
\[
\begin{array}{lll}
\hline
\nonum
\hskip 1cm h\in {\IQ}_8  &\hskip 2cm g\in SU(2)
&\hskip 2cm{\cal G}\in SO(3)
\nonum
\nonum
\hline
\nonum
%1%%%%%%%%%%%%%%%%%%%%%%%%%%%%%%%%%%%%%%%%%%%%%%%%%%%%%%%%%%%%%%%%%%
i=\pmatrix{i & 0\cr
	   0 & -i\cr}
		& g_i=\pmatrix{ e^{i(\theta_0+{\theta\ovr 4})} & 0 \cr
			      0  & e^{-i({\theta_0+{\theta\ovr 4}})}\cr}
				     & {\cal G}_i =
\pmatrix{ 1 &  0              &              0 \cr
 0 & \cos ({{\th\ovr 2}+2\th_0} )  & \sin ({{\th\ovr 2}+2\th_0} ) \cr
 0 & -\sin ({{\th\ovr 2}+2\th_0} ) & \cos ({{\th\ovr 2}+2\th_0} ) \cr  }
\nonum
\nonum
%2%%%%%%%%%%%%%%%%%%%%%%%%%%%%%%%%%%%%%%%%%%%%%%%%%%%%%%%%%%%%%%%%%%%
j=\pmatrix{ 0 & -1\cr
	    1 & 0 \cr}
		     & g_j
=\pmatrix{\cos(\theta_0+{\theta\ovr 4}) & -\sin(\theta_0+{\theta\ovr 4})\cr
	  \sin(\theta_0+{\theta\ovr 4}) &  \cos(\theta_0+{\theta\ovr 4})\cr}
				      & {\cal G}_j =
\pmatrix{\cos ({{\th\ovr 2}+2\th_0} ) &    0
				      & \sin ({{\th\ovr 2}+2\th_0} ) \cr
	       0       &    1     &        0      \cr
	-\sin ({{\th\ovr 2}+2\th_0} ) &    0
				      & \cos ({{\th\ovr 2}+2\th_0} ) \cr  }
\nonum
\nonum
%3%%%%%%%%%%%%%%%%%%%%%%%%%%%%%%%%%%%%%%%%%%%%%%%%%%%%%%%%%%%%%%%%%%%%%%%%
k=\pmatrix{ 0 & -i\cr
	   -i & 0 \cr}
		     & g_k
=\pmatrix{\cos(\theta_0+{\theta\ovr 4})
		& -i\,\sin(\theta_0+{\theta\ovr 4})\cr
       -i\,\sin(\theta_0+{\theta\ovr 4})
		&  \cos(\theta_0+{\theta\ovr 4})\cr}
				       & {\cal G}_k =
\pmatrix{ \cos ({{\th\ovr 2}+2\th_0} )
		& \sin ({{\th\ovr 2}+2\th_0} ) & 0 \cr
	 -\sin ({{\th\ovr 2}+2\th_0} )
		& \cos ({{\th\ovr 2}+2\th_0} ) & 0 \cr
		   0     &       0        & 1   \cr  }
\nonum
\nonum
%4%%%%%%%%%%%%%%%%%%%%%%%%%%%%%%%%%%%%%%%%%%%%%%%%%%%%%%%%%%%%%%%%%%%%%%%%
-i=\pmatrix{-i & 0\cr
	     0 & i\cr}
		     & g_i^{-1}=\pmatrix{ e^{-i(\theta_0+{\theta\ovr 4})}
						& 0 \cr
		     0  & e^{i({\theta_0+{\theta\ovr 4}})}\cr}
		     & {\cal G}_i^{-1} =
\pmatrix{ 1 &  0              &              0 \cr
	  0 & \cos ({{\th\ovr 2}+2\th_0} )
	  &-\sin ({{\th\ovr 2}+2\th_0} ) \cr
	  0 & \sin ({{\th\ovr 2}+2\th_0} )
	  & \cos ({{\th\ovr 2}+2\th_0} ) \cr}
\nonum
\nonum
\nonum
%5%%%%%%%%%%%%%%%%%%%%%%%%%%%%%%%%%%%%%%%%%%%%%%%%%%%%%%%%%%%%%%%%%%%%%%%
-j=\pmatrix{ 0 & 1\cr
	    -1 & 0 \cr}
		     & g_j^{-1}
 =\pmatrix{ \cos(\theta_0+{\theta\ovr 4})
		& \sin(\theta_0+{\theta\ovr 4})\cr
	   -\sin(\theta_0+{\theta\ovr 4})
		& \cos(\theta_0+{\theta\ovr 4})\cr}
				      & {\cal G}_j^{-1} =
\pmatrix{ \cos ({{\th\ovr 2}+2\th_0} ) &    0
		& -\sin ({{\th\ovr 2}+2\th_0} ) \cr
			  0       &    1     &        0      \cr
	  \sin ({{\th\ovr 2}+2\th_0} ) &    0
		& \cos ({{\th\ovr 2}+2\th_0} ) \cr  }
\nonum
\nonum
%6%%%%%%%%%%%%%%%%%%%%%%%%%%%%%%%%%%%%%%%%%%%%%%%%%%%%%%%%%%%%%%%%%%%%%%%
-k=\pmatrix{ 0 & i\cr
	     i & 0 \cr}
		     & g_k^{-1}
 =\pmatrix{ \cos(\theta_0+{\theta\ovr 4})
		& i\,\sin(\theta_0+{\theta\ovr 4})\cr
	    i\,\sin(\theta_0+{\theta\ovr 4})
		&  \cos(\theta_0+{\theta\ovr 4})\cr}
				       & {\cal G}_k^{-1} =
\pmatrix{ \cos ({{\th\ovr 2}+2\th_0} )
		& -\sin ({{\th\ovr 2}+2\th_0} ) & 0 \cr
	  \sin ({{\th\ovr 2}+2\th_0} )
		& \cos ({{\th\ovr 2}+2\th_0} ) & 0 \cr
		   0     &       0        & 1   \cr  }
\nonum
\nonum
%7%%%%%%%%%%%%%%%%%%%%%%%%%%%%%%%%%%%%%%%%%%%%%%%%%%%%%%%%%%%%%%%%%%%%%%%%
\I1=\pmatrix{1 & 0\cr
	     0 & 1\cr}
		     & g_\I1=\pmatrix{ 1  & 0 \cr
				       0  & 1 \cr}
				     & {\cal G}_\I1 =
\pmatrix{ 1 &  0  &  0 \cr
	  0 &  1  &  0 \cr
	  0 &  0  &  1 \cr}
\nonum
\nonum
%8%%%%%%%%%%%%%%%%%%%%%%%%%%%%%%%%%%%%%%%%%%%%%%%%%%%%%%%%%%%%%%%%%%%%%%
{-\I1}=\pmatrix{-1 &  0\cr
		 0 & -1\cr}
		     & g_{-\I1}=\pmatrix{ e^{i(\theta_0+{\theta\ovr 2})}
				& 0 \cr
				0  & e^{-i({\theta_0+{\theta\ovr 2}})}\cr}
				     & {\cal G}_{-\I1} =
\pmatrix{ 1 &  0              &              0 \cr
	  0 & \cos ({{\th}+2\th_0} )  & \sin ({{\th}+2\th_0} ) \cr
	  0 & -\sin ({{\th}+2\th_0} ) & \cos ({{\th}+2\th_0} ) \cr  }
\nonum
\nonum
%9%%%%%%%%%%%%%%%%%%%%%%%%%%%%%%%%%%%%%%%%%%%%%%%%%%%%%%%%%%%%%%%%%%%%%%
{-\I1}=\pmatrix{ -1 & 0 \cr
		  0 & -1\cr}
		     & g_{-\I1}
 =\pmatrix{ \cos(\theta_0+{\theta\ovr 2})
		& -\sin(\theta_0+{\theta\ovr 2})\cr
	    \sin(\theta_0+{\theta\ovr 2})
		&  \cos(\theta_0+{\theta\ovr 2})\cr}
				      & {\cal G}_{-\I1} =
\pmatrix{ \cos ({{\th}+2\th_0} ) &    0     & \sin ({{\th}+2\th_0} ) \cr
		0       &    1     &        0      \cr
	 -\sin ({{\th}+2\th_0} ) &    0     & \cos ({{\th}+2\th_0} ) \cr}
\nonum
\nonum
%10%%%%%%%%%%%%%%%%%%%%%%%%%%%%%%%%%%%%%%%%%%%%%%%%%%%%%%%%%%%%%%%%%%%%%%
-\I1=\pmatrix{ -1 & 0  \cr
		0 & -1 \cr}
		     & g_{-\I1}
 =\pmatrix{ \cos(\theta_0+{\theta\ovr 2})
		& -i\,\sin(\theta_0+{\theta\ovr 2})\cr
	    -i\,\sin(\theta_0+{\theta\ovr 2})
		&  \cos(\theta_0+{\theta\ovr 2})\cr}
				       & {\cal G}_{-\I1} =
\pmatrix{ \cos ({{\th}+2\th_0} ) & \sin ({{\th}+2\th_0} ) & 0 \cr
	 -\sin ({{\th}+2\th_0} ) & \cos ({{\th}+2\th_0} ) & 0 \cr
		   0     &       0        & 1   \cr  }
\nonum
\nonum
%%%%%%%%%%%%%%%%%%%%%%%%%%%%%%%%%%%%%%%%%%%%%%%%%%%%%%%%%%%%%%%%%%%%%%%%%%
\hline\end{array}
\] }
\caption{}
\end{table}
\renewcommand{\baselinestretch}{1.5}
{\footnotesize
\begin{table}
 \begin{center}
 \begin{tabular}{llcc}
 \hline
 \(\qquad\qquad U \quad\)&\(\qquad\tilde{{\cal G}_i}\qquad \) &
 \(\qquad\tilde{{\cal G}_j}\qquad \)&\( U<\Phi_{diag}>U^{-1} \)
 \\ \hline\\
 \(
 \pmatrix{-1 &  0 & 0 \cr
	   0 & -1 & 0 \cr
	   0 &  0 & 1 \cr  } \)
				&
				\(\qquad  {\cal G}_{-i} \)
						&
						\(  {\cal G}_{-j} \)
 & \( \pmatrix{\vf_1 &  0     & 0 \cr
		0    & \vf_2 & 0 \cr
		0    &  0     & -(\vf_1+\vf_2) \cr  } \)
						\\ \\
 \(
 \pmatrix{-1 &  0 & 0 \cr
	   0 &  1 & 0 \cr
	   0 &  0 &-1 \cr  } \)
				&
				\(\qquad   {\cal G}_{-i} \)
						&
						\(  {\cal G}_{j} \)
 & \( \pmatrix{\vf_1 &  0     & 0 \cr
		0    & \vf_2 & 0 \cr
		0    &  0     & -(\vf_1+\vf_2) \cr  } \)
						\\ \\
 \(
 \pmatrix{ 0 &  1 & 0 \cr
	   1 &  0 & 0 \cr
	   0 &  0 & 1 \cr  } \)
				&
				\(\qquad {\cal G}_{j} \)
						&
						\(  {\cal G}_{i} \)
 & \( \pmatrix{\vf_2 &  0     & 0 \cr
		0    & \vf_1 & 0 \cr
		0    &  0     & -(\vf_1+\vf_2) \cr  } \)
						\\ \\
 \hline
 \end{tabular}
 \end{center}
 \caption{}
 \label{tab2}
 \end{table}
}
%-------------------------Figure-------------------------------------
\clearpage

\samepage
\vskip 1cm
\begin{figure}[h]
	\epsfysize=230 pt
	\epsffile[40 250 300 550]{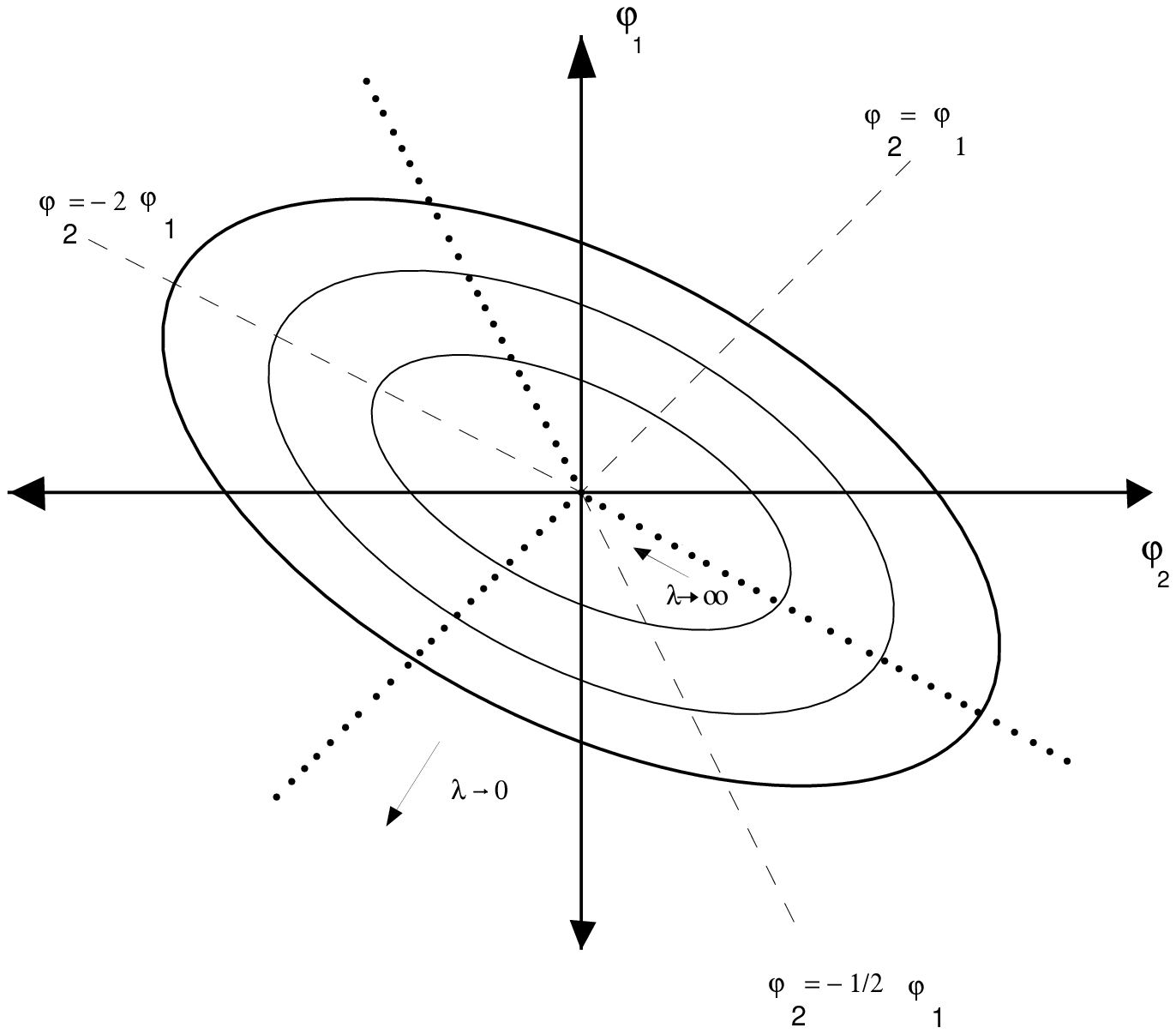}
       \caption{ }
       \label{ellipse4}
\end{figure}
\vskip 1 cm
\begin{figure}[h]
	\epsfxsize=300 pt
	\epsffile[20 300 400 600]{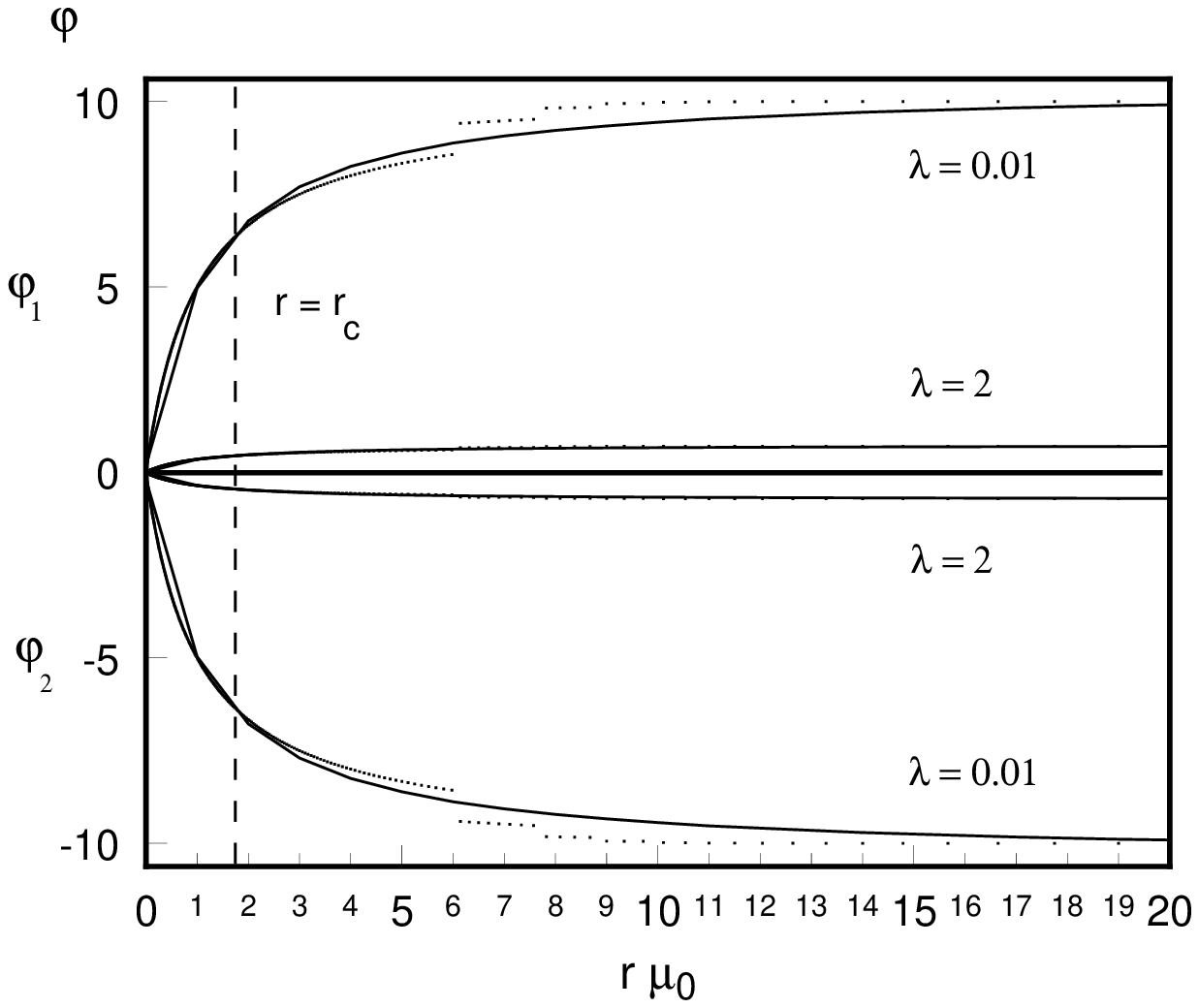}
	\caption{}
\end{figure}

\samepage
\vskip 1cm
\begin{figure}[h]
       \epsfysize=230 pt
       \epsffile[50 250 300 550 ]{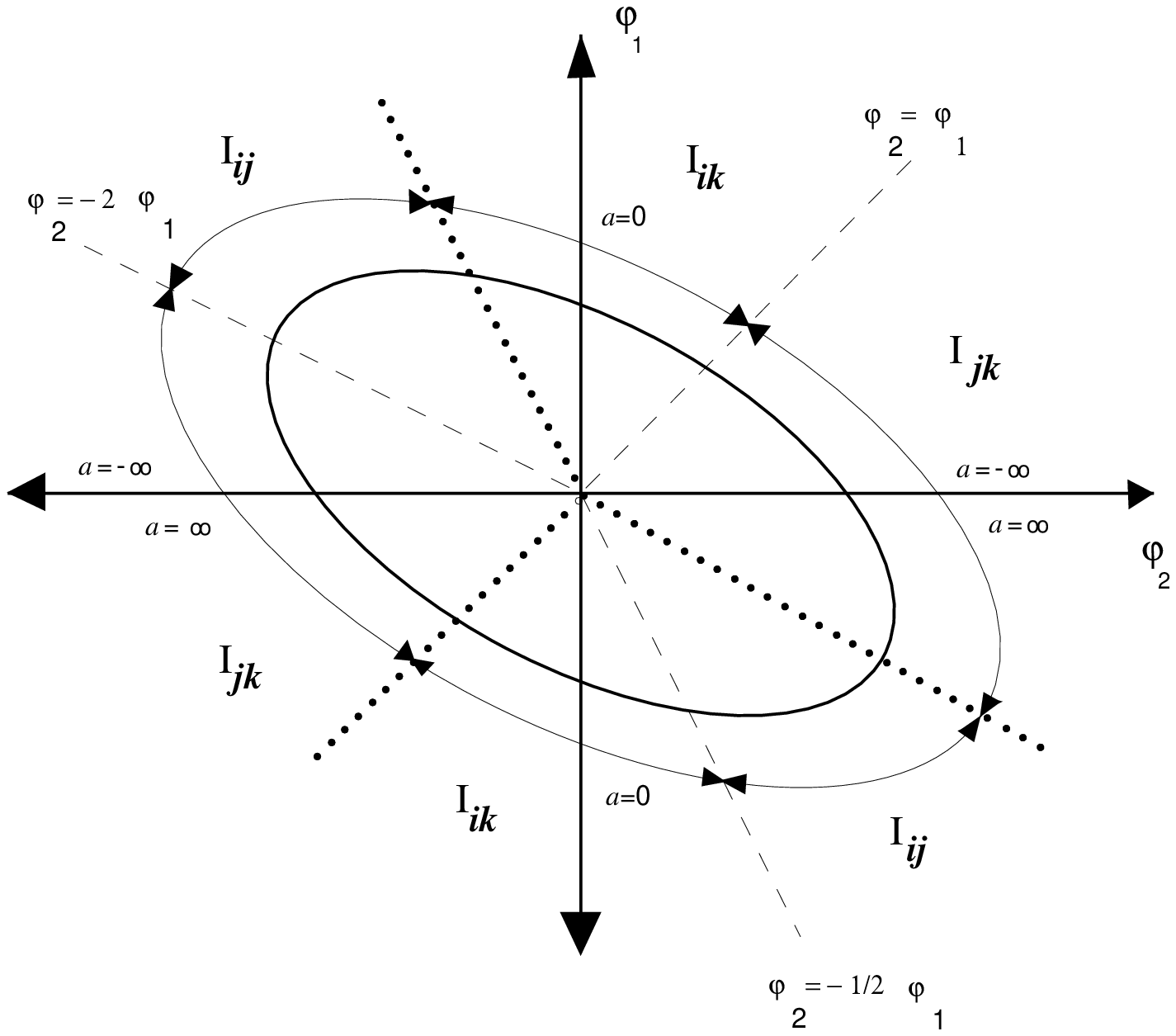}
       \caption{}
       \label{ellipse6}
\end{figure}
\begin{figure}[h]
	\epsfxsize=300 pt
	\epsffile[20 300 400 600]{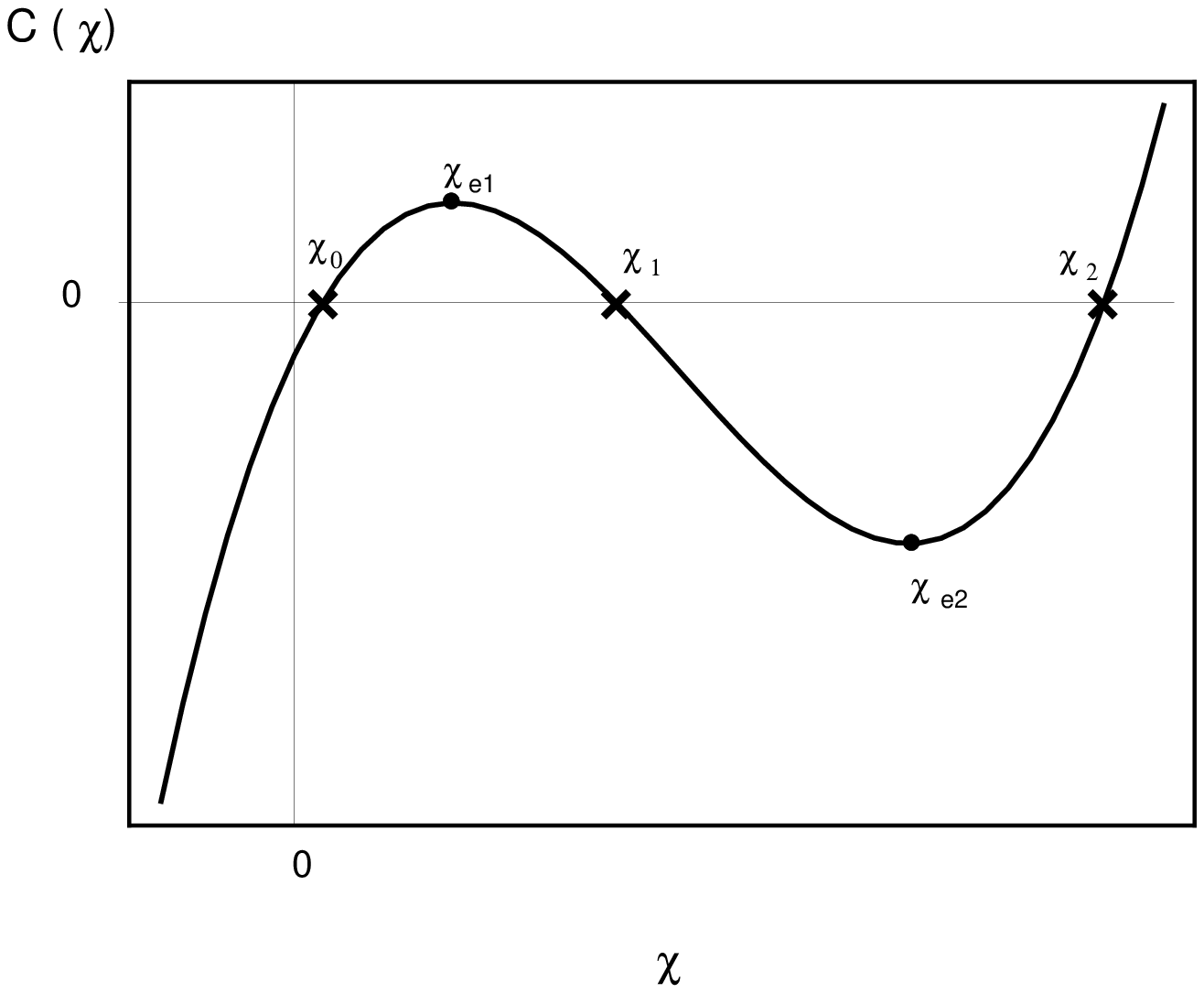}
	\caption{}
\end{figure}
\clearpage

\samepage
\vskip 1 cm
\begin{figure}[h]
	\epsfysize=220 pt
	\epsffile[0 220 450 550]{fea.ps}
	\begin{center} Figure 5a : \end{center}
\end{figure}
\vskip 1.5 cm
\begin{figure}[h]
	\epsfysize=220 pt
	\epsffile[0 220 450 550] {fabit.ps}
	\begin{center} Figure 5b : \end{center}
\end{figure}
\end{document}